\documentclass[journal=jctcce,manuscript=article]{achemso}
\usepackage[T1]{fontenc}
\usepackage{mathptmx}
\usepackage[scaled=0.92]{helvet}
\usepackage{amsmath,amssymb,graphicx,booktabs,array,tabularx,multirow}
\usepackage[hidelinks]{hyperref}
\setkeys{acs}{articletitle=true}
\newcommand{\Eh}{$E_\mathrm{h}$}
\newcommand{\mEh}{m$E_\mathrm{h}$}
\newcommand{\uEh}{$\mu E_\mathrm{h}$}
\newcommand{\kcal}{kcal\,mol$^{-1}$}
\newcommand{\Ms}{$M_S$}
\newcommand{\LiMnO}{Li$_{2-x}$MnO$_3$}
\newcommand{\figorbox}[3]{\IfFileExists{#2}{\includegraphics[width=#1]{#2}}%
  {\fbox{\parbox[c][#3][c]{\dimexpr#1-2\fboxsep-2\fboxrule\relax}{\centering\small Figure to be produced:\\ \texttt{\detokenize{#2}}}}}}

\title{Where Does the Anion-Redox Hole Go? Reliability of Open-Shell Projection-Based Embedding for Li-Rich Cathodes, from DFT Environments to Quantum Solvers}

\author{Pritam Das}
\email{pritam@kookmin.ac.kr}
\affiliation{School of Materials Science and Engineering, Kookmin University, Seoul, 02707, Republic of Korea}
\author{Satadeep Bhattacharjee}
\affiliation{Indo-Korea Science and Technology Center (IKST), Bengaluru 560064, India}
\author{Pil-Ryung Cha}
\affiliation{School of Materials Science and Engineering, Kookmin University, Seoul, 02707, Republic of Korea}
\email{cprdream@kookmin.ac.kr}

\abbreviations{DFT, DMRG, SQD, SPADE, IAO, NO, UNO}
\keywords{anion redox, projection-based embedding, DMRG, quantum computing, Li-rich cathodes}

\begin{document}

%\begin{tocentry}
%\centering
%\figorbox{\linewidth}{figures/toc.pdf}{3cm}
%\end{tocentry}

\begin{abstract}
Anion redox in Li-rich layered oxides places holes on oxygen next to an antiferromagnetic Mn$^{4+}$ (d$^3$) lattice, an open-shell, strongly correlated problem for which projection-based wavefunction-in-DFT embedding offers a route to correlated and quantum solvers, but whose error sources have been characterized mainly for closed-shell molecules. We develop an open-shell (spin-unrestricted) implementation of projection-based embedding, exact at the DFT-in-DFT level to the \uEh{} scale for closed-shell, broken-symmetry and $n_\alpha \neq n_\beta$ references with point charges and DFT+U, and apply DMRG-in-DFT to point-charge-embedded clusters of delithiated \LiMnO{}, varying the environment functional (seven, 0--100\,\% exact exchange), geometry, active region, active-space spin, virtual space, cluster size and magnetic reference. In every setting the lowest SCF state places the hole on an oxygen pair, and no O--O bond forms (Mayer bond order $\leq 0.24$). How the hole is shared is set by the functional together with the geometry: hybrid environments give a full, uncoupled hole at the DFT+U geometry, but a partial, weakly covalent and partly asymmetric hole at the HSE06 geometry, where the pair contracts from 2.58 to 2.41\,\AA{}; the change reaches 0.5\,e on one oxygen. For DFT+U and HSE06 at their own geometries, spin spreads across the active/environment boundary, so the integer spin projection of the active space becomes a choice worth up to 0.26\,e when the active region holds only the oxygen pair and $\leq 0.09$\,e in the region that adds its two Mn neighbors (R2). An extended state search found lower states, missed by the initial search, for two of the seven functionals, changing a pair-oxygen spin by up to 0.45\,e. Across all 82 production runs the embedded mean field already gives the same hole (DMRG changes a pair-oxygen spin by a median of 0.02--0.04\,e and never reverses it); what correlation mainly adds is O--O covalency. State control and geometry consistency, not the correlated solver, therefore dominate the uncertainty, and total energies, with them state rankings, are not converged at 30--40 correlated orbitals. For quantum solvers, the embedded Hamiltonian of a spin-polarized environment is intrinsically spin-unrestricted: a shared-orbital approximation reverses the hole, whereas sample-based quantum diagonalization with an unrestricted subspace solver reproduces exact diagonalization in 12- and 14-orbital reduced active spaces, as accurately as, but not more accurately than, a classical selected configuration interaction of the same size.
\end{abstract}

%=====================================================================
\section{Introduction}\label{sec:intro}
%=====================================================================

Li-rich layered oxides deliver capacities beyond the transition-metal redox limit because oxygen itself is oxidized on delithiation.\cite{sathiya2013,seo2016,luo2016} Li$_2$MnO$_3$ is the prototype: Mn is already Mn$^{4+}$ (d$^3$), so the extraction holes must go to O 2p. The oxidized oxygen may remain a localized O$^-$ hole, pair into O--O dimers (peroxo-like, $\sim$1.3\,\AA{} in hybrid-DFT calculations\cite{chen2016}), or delocalize over Mn--O states, and which of these happens controls voltage hysteresis and oxygen loss. Periodic DFT does not settle the question, because the hole character depends on the functional: semilocal and DFT+U functionals delocalize the hole, hybrid functionals localize it, and the answer changes with $U$ and with the exact-exchange fraction.\cite{seo2016,chen2016} The underlying physics is multireference, with a d$^3$ Mn$^{4+}$ sublattice coupled antiferromagnetically and O 2p holes whose spins couple to it.

Projection-based wavefunction-in-DFT (WF-in-DFT) embedding\cite{manby2012,lee2019} treats a small active region with a correlated solver, such as CCSD(T), CASSCF, DMRG or a quantum algorithm,\cite{ralli2024,rossmannek2021} and the rest of the system at the DFT level. It is formally exact at the DFT-in-DFT level and systematically improvable in the size of the active region, and it is a natural route to applying quantum solvers to materials: the active region is small enough for near-term devices, while the environment is handled classically. Recent work has shown where its accuracy is lost. Monino et al.\ found that the error of DMRG-in-DFT is dominated by the nonadditive exchange--correlation coupling between the active region and its environment, and that replacing the nonadditive DFT exchange by exact exchange removes most of it,\cite{monino2026jctc} building on earlier nonadditive corrections.\cite{goodpaster2014} The same authors showed that the approach cannot be exact even with the exact functional.\cite{monino2026jcp} Because the error grows with the density overlap across the boundary, the environment functional and the placement of the boundary\cite{welborn2018} are coupled error sources.

These analyses have been carried out mostly for closed-shell organic molecules. Open-shell embedding with the Huzinaga projector exists,\cite{graham2022,hegely2025} and spin-crossover energies of Fe complexes have been obtained with embedded CCSD(T).\cite{waldrop2025} Open-shell cathode materials, however, combine features that none of these studies faced: several antiferromagnetically coupled open-shell metal centers, oxygen holes whose location is itself the quantity of interest, an environment that is open-shell, and a point-charge representation of the crystal.

Battery materials are also a recurring target for quantum algorithms, from fault-tolerant estimates of cathode properties\cite{delgado2022,zini2023} and core-level spectroscopy\cite{fomichev2024,loaiza2026,kunitsa2025} to open-system models of cathode polarons.\cite{courtney2026} Closest to the present system, Kunitsa et al.\cite{kunitsa2025} simulated the oxygen K-edge electron energy loss spectrum of Li$_2$MnO$_3$ on a cluster model, with a fault-tolerant resource estimate for an oxygen-centered cluster with 18 active orbitals, and Loaiza et al.\cite{loaiza2026} estimated the resources for resonant inelastic X-ray scattering on Li-rich cathode clusters.
These studies target spectra and fault-tolerant resources of pristine models. None treats the ground-state hole of delithiated Li$_2$MnO$_3$ by embedding, or asks how such a result depends on the DFT starting point (Section~S1 of the Supporting Information records the literature search).

\begin{figure}[tbp]
\centering
\includegraphics[width=\textwidth]{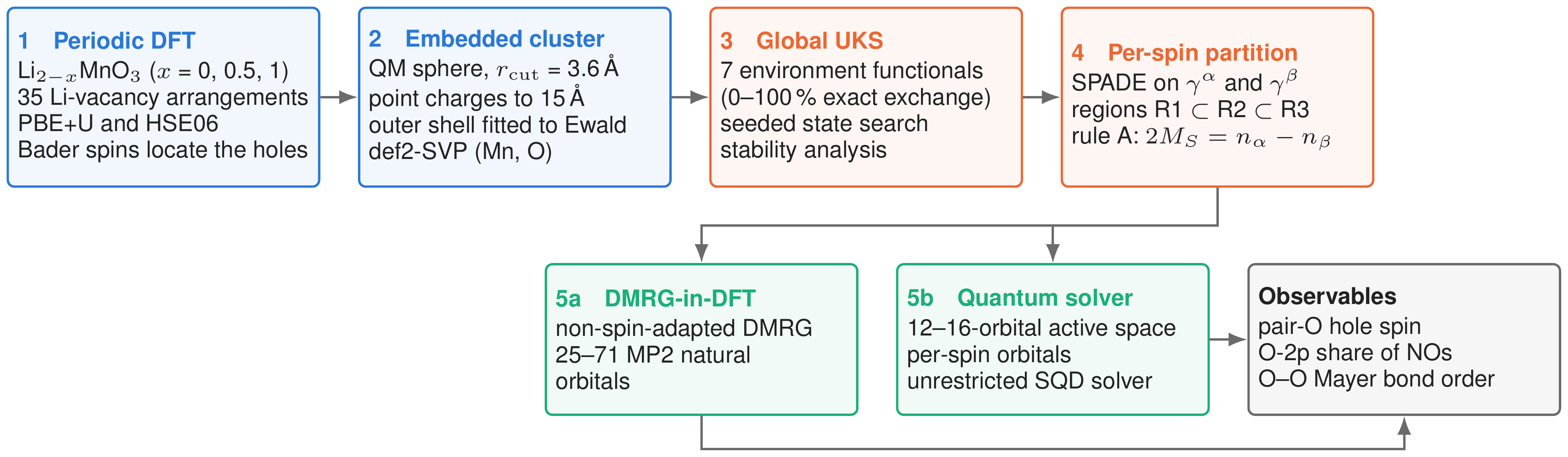}
\caption{Workflow of this study. Periodic calculations locate the holes and supply geometries and point charges (1); clusters are carved around the hole-bearing oxygen pair (2); a global unrestricted Kohn--Sham (UKS) calculation is converged for each of seven environment functionals with an explicit state search (3); the occupied space is partitioned per spin by SPADE into nested active regions, with the active-space spin projection fixed by rule A (4); the embedded problem is solved by non-spin-adapted DMRG with unrestricted orbitals (5a) or, in a device-sized active space, by sample-based quantum diagonalization (SQD) with an unrestricted subspace solver (5b).}
\label{fig:workflow}
\end{figure}

The environment functional that describes Mn--O bonding best (DFT+U, screened hybrids) is not the one that minimizes the embedding boundary error, which favors high exact exchange; high exact exchange, in turn, over-localizes transition-metal oxides. Here we ask how this trade-off decides where the anion-redox hole ends up (O 2p or Mn 3d, and with what O--O coupling) and how reliably it is described. We build and validate an open-shell (spin-unrestricted) projection-based embedding code (Section~\ref{sec:theory}) and apply it to point-charge-embedded clusters of delithiated \LiMnO{} ($x$ = 0, 0.5, 1) following the workflow of Figure~\ref{fig:workflow}. After validating the implementation against exactness tests and published closed- and open-shell benchmarks (Section~\ref{sec:validation}), we establish that the point-charge clusters reproduce the periodic hole (Section~\ref{sec:cluster}), and then vary the environment functional, the active region, the nonadditive exchange correction, the virtual space, the geometry and the magnetic reference one at a time, and test the completeness of the state search itself (Sections~\ref{sec:location}--\ref{sec:budget}). Finally, we solve the same embedded Hamiltonians with sample-based quantum diagonalization, separating the solver error, which a quantum device would have to beat, from the embedding error, which it cannot fix (Section~\ref{sec:sqd}).

%=====================================================================
\section{Theory}\label{sec:theory}
%=====================================================================

\subsection{Open-shell projection-based embedding}\label{sec:pbe}

A converged global Kohn--Sham (KS) calculation on the full system provides per-spin density matrices $\gamma^\sigma$ ($\sigma = \alpha, \beta$). The occupied orbitals of each spin are partitioned separately by the subsystem projected AO decomposition (SPADE)\cite{claudino2019spade} into an active set A and an environment set B,
\begin{equation}
\gamma^\sigma = \gamma^\sigma_\mathrm{A} + \gamma^\sigma_\mathrm{B}, \qquad \sigma = \alpha, \beta .
\label{eq:partition}
\end{equation}
Because the partition is carried out per spin, the active subsystem may hold different numbers of $\alpha$ and $\beta$ electrons and the two spin channels may have different environment orbitals; the restricted closed-shell case is the $\gamma^\alpha = \gamma^\beta$ instance of the same equations. The active subsystem is described in the field of the frozen environment through the embedded one-electron operator
\begin{equation}
h^\sigma_\mathrm{emb} = h + v^\sigma_\mathrm{emb} + P^\sigma, \qquad
v^\sigma_\mathrm{emb} = G^\sigma[\gamma^\alpha, \gamma^\beta] - G^\sigma[\gamma^\alpha_\mathrm{A}, \gamma^\beta_\mathrm{A}],
\label{eq:vemb}
\end{equation}
where $h$ contains the kinetic energy, the nuclear attraction and the external point charges, and $G^\sigma$ collects the Hartree potential of the total density and the exchange--correlation potential of the global functional, evaluated per spin (eq~7 of ref~\citenum{monino2026jctc}). Orthogonality to the environment is enforced either by the level-shift projection operator\cite{manby2012} or by the Huzinaga projection operator,\cite{hegely2016,graham2020}
\begin{equation}
P^\sigma_\mu = \mu\, S \gamma^\sigma_\mathrm{B} S, \qquad
P^\sigma_\mathrm{H} = -\left( F^\sigma \gamma^\sigma_\mathrm{B} S + S \gamma^\sigma_\mathrm{B} F^\sigma \right),
\label{eq:projectors}
\end{equation}
with $S$ the overlap matrix and $F^\sigma$ the Fock matrix, rebuilt at every SCF cycle. The total energy is
\begin{equation}
E^\text{WF-in-DFT} = E_\mathrm{WF}\!\left[\Psi_\mathrm{A}; h + v_\mathrm{emb}\right]
 + E_\mathrm{DFT}[\gamma] - E_\mathrm{DFT}[\gamma_\mathrm{A}]
 - \sum_\sigma \operatorname{tr}\!\left( \gamma^\sigma_\mathrm{A} v^\sigma_\mathrm{emb} \right),
\label{eq:energy}
\end{equation}
where $E_\mathrm{WF}$ is the energy of the correlated active-space wavefunction $\Psi_\mathrm{A}$ for the embedded Hamiltonian and the projector energy is not included. If the active subsystem is treated with the global functional, eq~\ref{eq:energy} reduces to $E_\mathrm{DFT}[\gamma]$, identically with the Huzinaga operator and in the limit $\mu \to \infty$ with the level shift; this DFT-in-DFT exactness is the central test of an implementation (Section~\ref{sec:validation}). We evaluate the energy both in the form of eq~\ref{eq:energy} (eq~1 of ref~\citenum{monino2026jctc}) and in their decomposed form (eq~18), which separates the nonadditive Hartree and exchange--correlation terms.

\subsection{Nonadditive exchange correction}\label{sec:dx}

The nonadditive exchange energy of the partition, $E_x^\mathrm{nadd} = E_x[\gamma] - E_x[\gamma_\mathrm{A}] - E_x[\gamma_\mathrm{B}]$, is described by the environment functional, partly or entirely by its semilocal part. Monino et al.\cite{monino2026jctc} replace it by its exact (Hartree--Fock) counterpart; we denote their nonadditive exchange correction $\Delta_x^\mathrm{nadd}$ by $\Delta x$,
\begin{equation}
\Delta x = E_x^\mathrm{nadd,HF} - E_x^\mathrm{nadd,DFT}, \qquad
E_x^\mathrm{nadd,HF} = -\sum_\sigma \operatorname{tr}\!\left[ \gamma^\sigma_\mathrm{A}\, K\!\left(\gamma^\sigma_\mathrm{B}\right) \right],
\label{eq:dx}
\end{equation}
where $K(\gamma)_{pq} = \sum_{rs} (pr|sq)\, \gamma_{rs}$ (eqs~20--21 of ref~\citenum{monino2026jctc}). Both terms are evaluated with the correlated active density, the one-particle reduced density matrix of the embedded wavefunction $\Psi_\mathrm{A}$ in the AO basis, and the frozen environment density $\gamma_\mathrm{B}$.
The spin-resolved form of eq~\ref{eq:dx} follows from the bilinearity of the exchange energy in the density, which we verified numerically. $\Delta x$ vanishes for a Hartree--Fock environment. It enters the energy only, not the embedding potential, so it cannot change the hole within a given state, but it can change which of two states is lower (Section~\ref{sec:energies}).

\subsection{Spin projection of the active space}\label{sec:rulea}

With a spin-unrestricted correlated solver, the per-spin numbers of active orbitals $n^\mathrm{A}_\alpha$ and $n^\mathrm{A}_\beta$ fix the spin projection of the active space,
\begin{equation}
2M_S = n^\mathrm{A}_\alpha - n^\mathrm{A}_\beta .
\label{eq:ms}
\end{equation}
For a closed-shell or fully localized open-shell subsystem this number is unambiguous. When orbitals are shared across the boundary, however, SPADE's criterion, the largest gap in the singular-value spectrum, becomes ambiguous (singular values of about 0.5--0.8), and nearly identical singular-value spectra can lead to different cuts. We therefore tie the orbital counts to the state of the global SCF. Let
\begin{equation}
p = \sum_{I \in \mathrm{A}} \left( q^\alpha_I - q^\beta_I \right)
\label{eq:p}
\end{equation}
be the spin population of the active atoms in intrinsic atomic orbitals (IAOs).\cite{knizia2013} If $p$ lies within 0.25 of an integer, the counts are set so that $2M_S$ equals that integer, with the smallest change from SPADE's cut. Otherwise the spin is spread across the boundary, both bracketing values $2M_S = \lfloor p \rfloor$ and $\lceil p \rceil$ (the two spin-count options) are computed, and every result is reported as the interval between them. A mean of two spin sectors is not a physical spin and is never formed. We refer to this protocol as rule A.

\subsection{Spin-unrestricted embedded Hamiltonian for quantum solvers}\label{sec:uham}

A quantum or selected-CI solver acts on a reduced active space of orbitals $\{\phi^\sigma_p\}$ inside the active space. Each spin's reduced active space must be orthogonal to that spin's environment, so when $\gamma^\alpha_\mathrm{B} \neq \gamma^\beta_\mathrm{B}$ the $\alpha$ and $\beta$ reduced-active-space orbitals differ in general. The embedded Hamiltonian in the reduced active space is then
\begin{equation}
\hat H = E_0 + \sum_\sigma \sum_{pq} h^\sigma_{pq}\, \hat a^\dagger_{p\sigma} \hat a_{q\sigma}
 + \frac12 \sum_{\sigma\sigma'} \sum_{pqrs} \left( p^\sigma q^\sigma \middle| r^{\sigma'} s^{\sigma'} \right)
 \hat a^\dagger_{p\sigma} \hat a^\dagger_{r\sigma'} \hat a_{s\sigma'} \hat a_{q\sigma},
\label{eq:uham}
\end{equation}
where $h^\sigma$ contains the embedded operator of eq~\ref{eq:vemb} and the mean field of the frozen orbitals outside the reduced active space, and $E_0$ is the corresponding constant. Equation~\ref{eq:uham} has two one-body matrices and three distinct two-electron blocks, $(\alpha\alpha|\alpha\alpha)$, $(\alpha\alpha|\beta\beta)$ and $(\beta\beta|\beta\beta)$. Under the Jordan--Wigner mapping $\alpha$ and $\beta$ spin orbitals are assigned separate qubits, so eq~\ref{eq:uham} is an ordinary qubit Hamiltonian. In sample-based quantum diagonalization (SQD),\cite{sqd2025} which extends quantum-selected configuration interaction\cite{kanno2023} with configuration recovery, bitstrings sampled from a quantum circuit are corrected by configuration recovery, and the Hamiltonian is projected onto and diagonalized in the space of determinants formed from the recovered $\alpha$ and $\beta$ strings. Current SQD software performs this classical step for a single one-body matrix and a single two-electron tensor; Section~\ref{sec:sqd} shows the consequences of imposing that restriction on eq~\ref{eq:uham} and the remedy.

%=====================================================================
\section{Computational Details}\label{sec:compdet}
%=====================================================================

\subsection{Implementation and molecular benchmarks}\label{sec:impl}

The embedding was implemented on top of PySCF 2.14.0\cite{pyscf} (default integration grids, level~3; the embedded SCF reuses the global grid; density fitting for the cathode clusters), with block2 0.5.3\cite{block2} for DMRG; the code is described in Section~S7 of the Supporting Information. The Huzinaga projection operator, or the level shift with $\mu = 10^8$, is used wherever \uEh{} fidelity matters (Section~\ref{sec:validation}). Embedded FCI is solved for several roots and the lowest root, or the lowest with the requested $\langle S^2 \rangle$, is selected.

The molecular benchmarks are water along one O--H bond (ten geometries, 0.4--5.0\,\AA{}; STO-3G, 6-31G and cc-pVDZ), MnO (sextet), the Mn(IV) d$^3$ oxo-aqua model [MnO(H$_2$O)]$^{2+}$ (quartet), MnO$_4^-$, [Fe(H$_2$O)$_6$]$^{3+}$ (S = 5/2 vs 3/2, def2-TZVP, Fe active, SCAN environment), and the C$\equiv$N dissociation of propionitrile (cc-pVDZ, B3LYP environment, $-$CN active, first concentric localization (CL) shell, $n = 0$,\cite{claudino2019cl} 35 orbitals) of Monino et al.\cite{monino2026jctc} Broken-symmetry references of water are obtained from a fragment guess (H$\alpha$ + OH with swapped spins).

\subsection{Periodic calculations}\label{sec:periodic}

Periodic calculations used VASP 6.3.2\cite{vasp1996} with PAW potentials\cite{blochl1994,kresse1999} (Li\_sv, Mn\_pv, O), a 520\,eV plane-wave cutoff, and spin-polarized PBE+U in the rotationally invariant form of Dudarev et al.\cite{dudarev1998} with $U$ = 3.9\,eV on Mn 3d, the value used by the Materials Project.\cite{jain2011} Li$_2$MnO$_3$ ($C2/m$) is modeled in its conventional cell of four formula units (Li$_8$Mn$_4$O$_{12}$), and every cell was started from N\'eel-type antiferromagnetic order within the Mn honeycomb layer.\cite{balamurugan2014} The pristine and $x$ = 0.5 cells keep this order; in the two lowest $x$ = 1 cells (cfg07 and cfg13) one of the four Mn reverses during the PBE+U relaxation, giving three parallel and one antiparallel Mn moment per cell, and HSE06 keeps this pattern. The embedded clusters inherit the Mn pattern of their periodic cell. The cell holds a single Mn layer, so successive layers are stacked ferromagnetically ($\mathbf{k} = 0$), whereas experiment finds antiferromagnetic interlayer stacking, $\mathbf{k} = (0, 0, \tfrac12)$;\cite{balamurugan2014} the Mn layers are 4.8\,\AA{} apart.
 For $x$ = 0.5 and 1 all 35 symmetry-inequivalent Li/vacancy arrangements were relaxed (cell and ions) on a $\Gamma$-centered $6\times4\times6$ mesh with Gaussian smearing of 0.05\,eV and electronic and ionic thresholds of $10^{-6}$\,eV and 0.01\,eV\,\AA$^{-1}$; the O ions were seeded with a small moment so that O 2p holes could polarize. HSE06\cite{hse06} single points at the PBE+U geometries used the same cutoff and a $\Gamma$-centered $4\times3\times4$ mesh for the density and the exact exchange, starting from the PBE wavefunctions. The two lowest $x$ = 1 cells were also relaxed with HSE06 at fixed cell ($3\times2\times3$ mesh, force threshold 0.02\,eV\,\AA$^{-1}$), with a PBE+U relaxation on the same mesh as control; cfg07 converged to 0.018\,eV\,\AA$^{-1}$, whereas cfg13 stopped at 0.036\,eV\,\AA$^{-1}$, with energy changes below 0.1\,meV over its last steps. Both relaxations ran in restart segments; energies are summed over all segments and given per cell (four formula units). Atomic charges and spins are obtained by Bader partitioning of the charge and magnetization densities.\cite{henkelman2006}

\subsection{Embedded clusters and active regions}\label{sec:clusters}

\begin{figure}[tbp]
\centering
\figorbox{\textwidth}{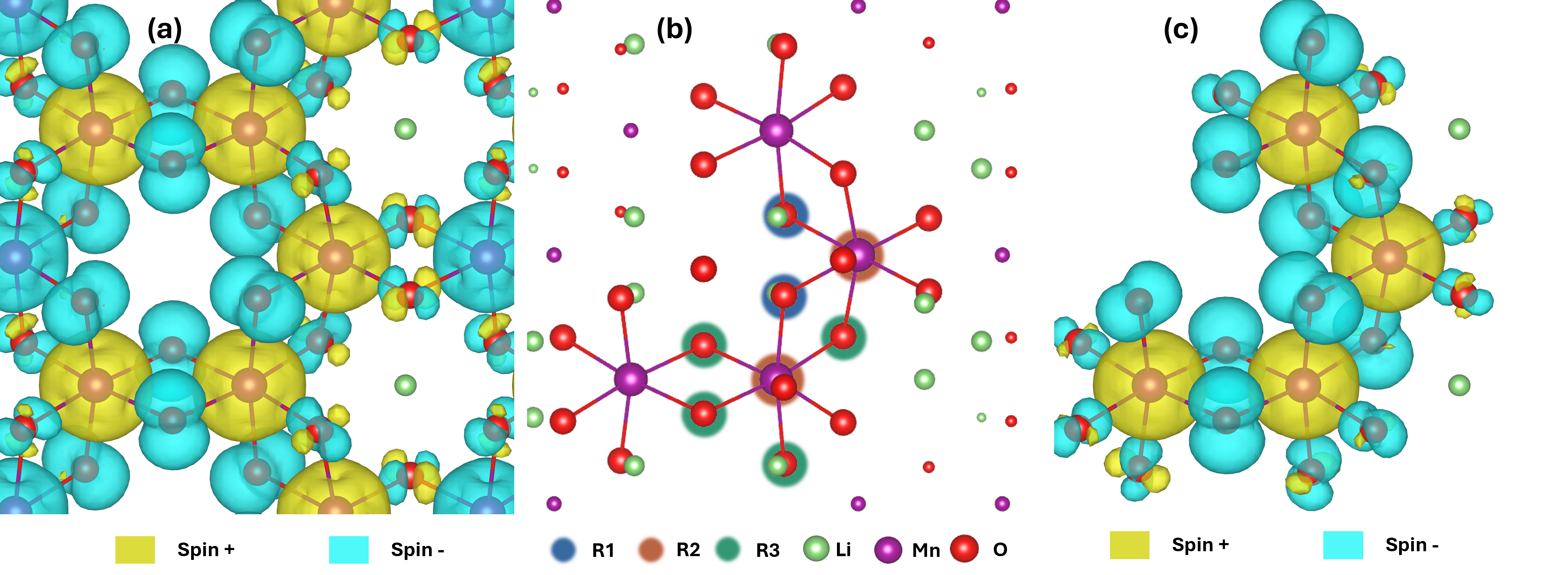}{6cm}
\caption{Cluster model of delithiated Li$_2$MnO$_3$ ($x$ = 1, cfg07, PBE+U geometry, $r_\mathrm{cut}$ = 3.6\,\AA{}), viewed along the normal of the Mn layer that contains the target O pair; all panels share the view. (a) Periodic HSE06 magnetization density of that MnO$_2$ layer (slab of $\pm$2.3\,\AA{}). (b) QM cluster in its point-charge field (point charges within 3\,\AA{} of the layer as small spheres), with the nested active regions R1 (O pair), R2 (+ two Mn) and R3 (+ neighboring hole-bearing and bridging O). (c) Spin density of the embedded DMRG-in-DFT solution in R2 (HSE06 environment, state a, the lower state of the extended search, NO threshold 0.002, spin counts 32/28), environment and active densities combined, in the same slab. Isosurfaces at $\pm$0.01\,e\,bohr$^{-3}$; yellow positive, cyan negative (the hole spin at $x$ = 1 is negative); the blue R1 halos in (b) mark the pair oxygens.}
\label{fig:cluster}
\end{figure}

Clusters are carved around the target oxygen of the dominant hole-bearing pair (Figure~\ref{fig:cluster}). The quantum-mechanical (QM) part contains all atoms within a radius $r_\mathrm{cut}$, completed to full MnO$_6$ octahedra, together with every Li bonded to a cluster O; the resulting sizes are listed in Table~\ref{tab:clusters}. The rest of the crystal, out to 15\,\AA{}, is represented by formal point charges (Li $+1$, Mn $+4$, O $-2$ plus its periodic hole occupation). The outermost 3\,\AA{} shell of charges is refitted by least squares so that the external potential at every QM site reproduces the periodic Ewald potential, following the approach of Klintenberg et al.\cite{klintenberg2000} (residual rms 18--24\,mV), with the total charge constrained to be exactly neutral. The cluster calculations use def2-SVP\cite{weigend2005} on Mn and O and STO-3G on Li$^+$, with density fitting. DFT+U uses PySCF's projection onto minimal-basis (MINAO) Mn 3d orbitals with $U$ = 3.0, 3.9 or 5.0\,eV. The broken-symmetry state is prepared from the ferromagnetic solution by exchanging the $\alpha$ and $\beta$ blocks of the Mn that are spin-down in the periodic calculation.

\begin{table}[tbp]
\caption{Number of QM atoms per cluster.}
\label{tab:clusters}
\centering
\begin{tabular}{lccc}
\toprule
cell & $r_\mathrm{cut}$ = 2.3\,\AA{} & 3.6\,\AA{} & 4.5\,\AA{} \\
\midrule
$x$ = 0.5, cfg04 & 28 & 54 & 87 \\
$x$ = 1, cfg07   & 22 & 44 & 71 \\
$x$ = 1, cfg13   & -- & 38 & 74 \\
\bottomrule
\end{tabular}
\end{table}

Three nested active regions are defined from the periodic Bader spins, in cluster coordinates with periodic images resolved, and are fixed before any embedded calculation: R1 is the target oxygen and its nearest hole-bearing oxygen (the pair and dimerization candidate); R2 adds the two Mn bonded to it; R3 adds the second bridging oxygen of the Mn--Mn edge and every hole-bearing oxygen bonded to an R2 Mn. In the production cell ($x$ = 1, cfg07) the pair is 2.58\,\AA{} apart and R1, R2 and R3 contain 2, 4 and 8 atoms. At $x$ = 0.5 (cfg04) the partner oxygen (2.71\,\AA{}) sits cis on one Mn octahedron rather than on the shared Mn--Mn edge; in cfg13 the pair is the shared edge itself (2.48\,\AA{}, the shortest O--O distance in the cluster), and R3 contains both hole pairs.

\subsection{Correlated calculations and observables}\label{sec:corr}

\paragraph{State tracking.} Every cluster SCF is started from the ferromagnetic solution, its periodic-pattern spin flip, single-Mn flips and cross-seeds from the other functionals' solutions. Only SCF failures are rejected. Every candidate is stability-tested (PySCF internal stability analysis, and a frontier-rotation test for DFT+U) and classified by hole location, Mn spin pattern and hole-spin orientation. We report the lowest stable state (state a) and, where it differs, the O-pair state with the periodic Mn pattern. Unstable candidates are followed rather than discarded. After the production runs the search was extended by two further routes (Section~\ref{sec:location}): an ECP anneal, in which the SCF is first converged with Stuttgart small-core effective core potentials on the Mn and Li point charges within 4\,\AA{} of the QM region and then reconverged without them, and cross-seeding: at the PBE+U geometry every functional was seeded from the lower HSE06 state found by the anneal, at the HSE06 geometry each hybrid was seeded from the HF state, from the other hybrid's state and from the lower states found at the PBE+U geometry, and at $x$ = 0.5 the periodic-pattern hybrid states were seeded from the HF and PBE+U states. Where a lower stable state was found, it replaces the initially recorded state, which we denote state a$'$.

\paragraph{DMRG.} The correlated solver is non-spin-adapted (SZ) DMRG\cite{block2} on the embedded UHF orbitals, which conserves only \Ms{} (set by rule A, Section~\ref{sec:rulea}). Each run is started twice, from the Hartree--Fock determinant and from an initial-guess matrix product state taken from a three-root state-averaged run, and $\langle S^2 \rangle$ is evaluated from the reduced density matrices. All runs use the same bond-dimension schedule, $M$ = 500, 750, 1000, 1500, 2000, trimmed to what fits in memory, with a checkpoint after each $M$; the change between the two largest $M$ is taken as the error bar. Memory limits R3 to $M$ = 750. The virtual space consists of MP2 natural orbitals\cite{taube2005} with occupations between 0.002 and 0.998 (25--37 orbitals in R2, 68--71 in R3), with thresholds of 0.001 and 0.0005 as convergence checks. For every production run the embedded UHF solution (HF-in-DFT), the reference from which the DMRG starts, is also evaluated with the same partition, spin counts and observables (82 runs), to separate what the correlated solver adds from what the embedding already fixes.

\paragraph{Observables.} The location of the hole is the lowest stable state, with the hole on the O pair, on Mn or on other oxygens; it is decided at the embedded-cluster DFT level for all functionals and checked at the correlated level. Within one state, the hole character is measured by (i) the spin on the two pair oxygens (pair-O spin), in symmetric L\"owdin ($S^{1/2}\gamma S^{1/2}$) and IAO\cite{knizia2013} populations of the total density; (ii) the O-2p share of the fractionally occupied natural orbitals of the correlated active density, in which each natural orbital is weighted by its distance from the nearest integer occupation, so that closed shells and the singly occupied Mn-3d moments drop out (meaningful only when the region contains Mn, R2 and R3); and (iii) the O--O Mayer bond order.\cite{mayer1983} Pair-O spins are reported with the sign of the state (negative for the hole states at $x$ = 1); figures plot the hole spin, the negative of the pair-O spin. As a negative control, the $x$ = 0 cluster gives pair spins of 0.014 and 0.001, as expected without a hole.

\subsection{Sample-based quantum diagonalization}\label{sec:sqddet}

The SQD tests use the HSE06 R2 problem at cfg07 (PBE+U geometry) in state a$'$, the state recorded before the extended search (Section~\ref{sec:location}), with spin counts 32/28. Reduced active spaces of 12, 14 and 16 orbitals (24--32 qubits) are built from the unrestricted natural orbitals (UNOs, the natural orbitals of the spin-summed embedded UHF density),\cite{pulay1988} and each spin's reduced active space is projected off its own environment and L\"owdin-orthonormalized, staying closest to the UNOs (Section~\ref{sec:sqd}). Bitstrings ($10^5$) are sampled from a unitary cluster Jastrow (UCJ) circuit\cite{lucj2023} built with ffsim\cite{ffsim} from UCCSD amplitudes, in its spin-unbalanced form (\texttt{UCJOpSpinUnbalanced}, in which the $\alpha\alpha$ and $\beta\beta$ Jastrow parameters may differ). To mimic hardware noise, a fraction $p$ = 0, 0.5 or 0.9 of the shots is replaced by uniformly random bitstrings. Configuration recovery and batching (300, 1000 or 3000 samples per batch) use qiskit-addon-sqd;\cite{qiskitsqd} its subspace diagonalization is replaced, through the \texttt{sci\_solver} hook, by a solver for eq~\ref{eq:uham} that applies the full unrestricted Hamiltonian restricted to the sampled determinant space with a Davidson eigensolver on a masked full-space contraction (about 30 lines; Section~S7 of the Supporting Information). Exact diagonalization (FCI) of each reduced active space is the reference. As classical baselines, the same Hamiltonian is diagonalized, with the same solver and at the same subspace dimension, in product subspaces of $\alpha$ and $\beta$ strings chosen (i) at random (the reference strings always included; three seeds), (ii) by their marginal weight in the FCI vector, and (iii) by a classical selected configuration interaction grown iteratively from the reference determinant, which adds the strings with the largest first-order coupling $|\langle D|\hat H|\Psi\rangle|$ to the current wavefunction (the CIPSI and heat-bath criterion).

%=====================================================================
\section{Results and Discussion}\label{sec:results}
%=====================================================================

\subsection{Validation of the open-shell implementation}\label{sec:validation}

\paragraph{DFT-in-DFT exactness.} For water along the O--H dissociation coordinate with B3LYP5, PBE0 and Hartree--Fock environments, the Huzinaga projector reproduces the global energy to better than 0.001\,\uEh{} at every geometry. The level-shift projector is exact only as $\mu \to \infty$: its residual scales as $1/\mu$ ($-17.5$, $-1.75$, $-0.175$ and $-0.0175$\,\uEh{} at $\mu$ = $10^5$, $10^6$, $10^7$ and $10^8$), because a finite shift lets the embedded orbitals acquire an $O(1/\mu)$ admixture of the environment space. The commonly used $\mu = 10^6$ therefore carries an intrinsic error of about 2\,\uEh{}. With all atoms active, the embedded calculation reduces to full-system FCI and reproduces it to 1\,\uEh{} state by state; the restricted FCIDUMP export round-trips to 1\,\uEh{}; and the two energy forms of Section~\ref{sec:pbe} agree to 1\,\uEh{}.

\paragraph{Broken-symmetry references.} Beyond $r \approx$ 1.3--1.5\,\AA{}, every restricted reference we examined (HF, BHandHLYP, PBE0, B3LYP, B3LYP5, $\omega$B97X) is externally unstable toward a spin-unrestricted, broken-symmetry\cite{noodleman1981} solution, so the restricted embedding references commonly used for bond dissociation are saddle points there. The fully broken-symmetry solution ($\langle S^2 \rangle \approx 1.00$ at 5\,\AA{}) is not reached by continuation from the restricted branch, which stalls on a partially broken branch ($\langle S^2 \rangle \approx 0.76$) lying 67--112\,\mEh{} higher, but is obtained from a fragment guess. For these spin-polarized references ($n_\alpha = n_\beta$, $\alpha \neq \beta$ orbitals) the DFT-in-DFT energy reproduces the global UKS energy to $\leq$1\,\uEh{} at all ten geometries for B3LYP5, PBE0, HF, PBE and SCAN, with both projectors. An existing implementation (Nbed 0.0.10\cite{ralli2024}) deviates by up to 0.15\,\Eh{} in this case, because it evaluates the exchange--correlation energy of the total density from a spin-summed density matrix (Section~S2 of the Supporting Information; reported as Nbed issue \#198); with that corrected, it agrees with our FCI-in-DFT values.

\paragraph{Open shells with $n_\alpha \neq n_\beta$.} For MnO (sextet; $n_\mathrm{act}$ = 14$\alpha$/9$\beta$ and 15$\alpha$/9$\beta$) and [MnO(H$_2$O)]$^{2+}$ (quartet; 17$\alpha$/14$\beta$) with PBE and B3LYP5 environments, the DFT-in-DFT energy reproduces the global UKS energy to $\leq$0.002\,\uEh{}. For the closed-shell d$^0$ anion MnO$_4^-$, HF-in-DFT agrees with the reference implementation to 5\,\uEh{}. Two practical observations matter for the cathode. SPADE applied per spin can place different numbers of environment orbitals in the two spin channels (MnO/B3LYP5: $n_\mathrm{env}$ = 4$\alpha$/5$\beta$), which a correlated solver must accommodate. And the UKS references are spin-contaminated ($\langle S^2 \rangle$ = 4.56 vs 3.75 for the Mn(IV) quartet with B3LYP5), which is why the non-spin-adapted DMRG solver was benchmarked against spin-adapted DMRG. On the MnO sextet (def2-SVP, all atoms active, 45 orbitals), non-spin-adapted (SZ) DMRG on UHF orbitals lies above spin-adapted (SU(2)) DMRG on ROHF orbitals at every bond dimension, by 11.6, 4.0 and 2.4\,\mEh{} at $M$ = 500, 1000 and 1500, and the two extrapolated limits agree within their extrapolation uncertainty; $\langle S^2 \rangle$ from the SZ density matrices is 8.7515--8.7575 against 8.75 for a pure sextet. With a frozen core the two differ by 1.9\,\mEh{}, because UHF and ROHF cores define different frozen-core problems; embedded and full-system calculations are therefore always compared with the same frozen-core convention.

\paragraph{Published closed-shell results.} We reproduce the FCI-in-DFT water dissociation of Ralli et al.\cite{ralli2024} directly against their Tables~V and VI. Table~VI is reproduced to $\leq$3.6\,\uEh{} for $r \leq 3$\,\AA{}; at 4--5\,\AA{} the deviation grows to $\sim$2\,\mEh{}, where the published values are themselves non-monotonic. The dissociation error derived from their Table~V (B3LYP5 environment) is 19.41\,\kcal{} against our 19.44. With both projectors and three environment functionals, our energies also agree with their implementation (Nbed 0.0.10 with the correction of Section~S2) to within 5\,\uEh{} for $r < 3$\,\AA{}, once both codes start from the same global SCF solution. Two systematic differences were identified rather than absorbed into tolerances. First, the reference implementation includes the projector energy $\operatorname{tr}[\gamma \mu P]$ in the total energy, whereas eq~\ref{eq:energy} does not; the conventions differ by $O(1/\mu)$ (0.36\,\uEh{} at $\mu = 10^8$ for the most compressed geometry). Second, at stretched geometries ($r \geq 3$\,\AA{}), where singlet and triplet states of the embedded fragment are nearly degenerate, a single-root Davidson solver can converge to the triplet or to an excited singlet, missing the singlet ground state by up to 190\,\uEh{} in a test where the embedded Hamiltonian is the full-system one; with several roots our values are never above the reference implementation's. For the propionitrile C$\equiv$N dissociation of Monino et al.,\cite{monino2026jctc} the global KS energies agree to $\leq$0.02\,\mEh{} and CAS(6,6)-in-B3LYP to 0.04--0.23\,\mEh{} for $r \leq 2.20$\,\AA{}, a constant offset consistent with different integration grids. At 2.35 and 2.50\,\AA{} CAS(6,6) has several local minima, and our lowest solutions lie 5.0 and 25.5\,\mEh{} below the published energies, so these points are not used. DMRG-in-B3LYP\cite{beran2023} gives a dissociation error at 2.50\,\AA{} of 24.75\,\kcal{} (published 27.31), and the nonadditive exchange correction lowers it by 10.5\,\kcal{} (published $\sim$10). The effect of the correction is therefore reproduced; the remaining 2.6\,\kcal{} difference in the uncorrected error (4.1\,\mEh{}) is not resolved, and the projector (ref~\citenum{monino2026jctc} used the parameter-free projection of Khait and Hoffmann\cite{khait2012} for DMRG-in-DFT), DMRG convergence or their self-consistent SCF-DMRG-in-DFT variant could each account for it.

\paragraph{Published open-shell results.} Table~\ref{tab:fe} compares the embedded spin-crossover energy of [Fe(H$_2$O)$_6$]$^{3+}$ with that of Waldrop et al.\cite{waldrop2025} The published values are recovered with the conventions of that work: a Huzinaga projector built from the full-system Fock matrix, the projector energy kept, and the level-shifted environment orbitals, which hold the environment electrons, kept in the correlation space. Freezing them, the partition-consistent choice used throughout this work, moves the CCSD(T)-in-SCAN gap by 4.4\,\kcal{}, to $-47.7$\,\kcal{} (MP2: by 2.7\,\kcal{}, or 2.3\,\kcal{} with a dynamic projector), close to the reference value back-corrected from experiment, $-47.3$\,\kcal{}. The correlation-space convention therefore changes embedded spin gaps by several \kcal{} and should be stated with them. For [Fe(H$_2$O)$_6$]$^{2+}$ our lowest high-spin SCAN solution lies 5.6\,\mEh{} above the published one (a different, quasi-degenerate d occupation), so its gap was not compared.

\begin{table}[tbp]
\caption{High-Spin--Low-Spin Gap of [Fe(H$_2$O)$_6$]$^{3+}$ (\kcal{})\textsuperscript{\emph{a}}}
\label{tab:fe}
\centering
\begin{tabular}{p{6.2cm}ccc}
\toprule
method & this work & ref~\citenum{waldrop2025} & difference \\
\midrule
SCAN & $-33.71$ & $-33.67$ & $-0.04$ \\
MP2-in-SCAN & $-46.09$ & $-46.40$ & $+0.31$ \\
CCSD(T)-in-SCAN & $-43.25$ & $-43.33$ & $+0.07$ \\
CCSD(T)-in-SCAN, environment orbitals frozen & $-47.74$ & -- & -- \\
\bottomrule
\end{tabular}

\raggedright\footnotesize\textsuperscript{\emph{a}}S = 5/2 vs 3/2, def2-TZVP, Fe active. Reference value back-corrected from experiment: $-47.3$\,\kcal{}.
\end{table}

\subsection{What controls the embedding error: water}\label{sec:water}

\begin{figure}[tbp]
\centering
\includegraphics[width=0.5\textwidth]{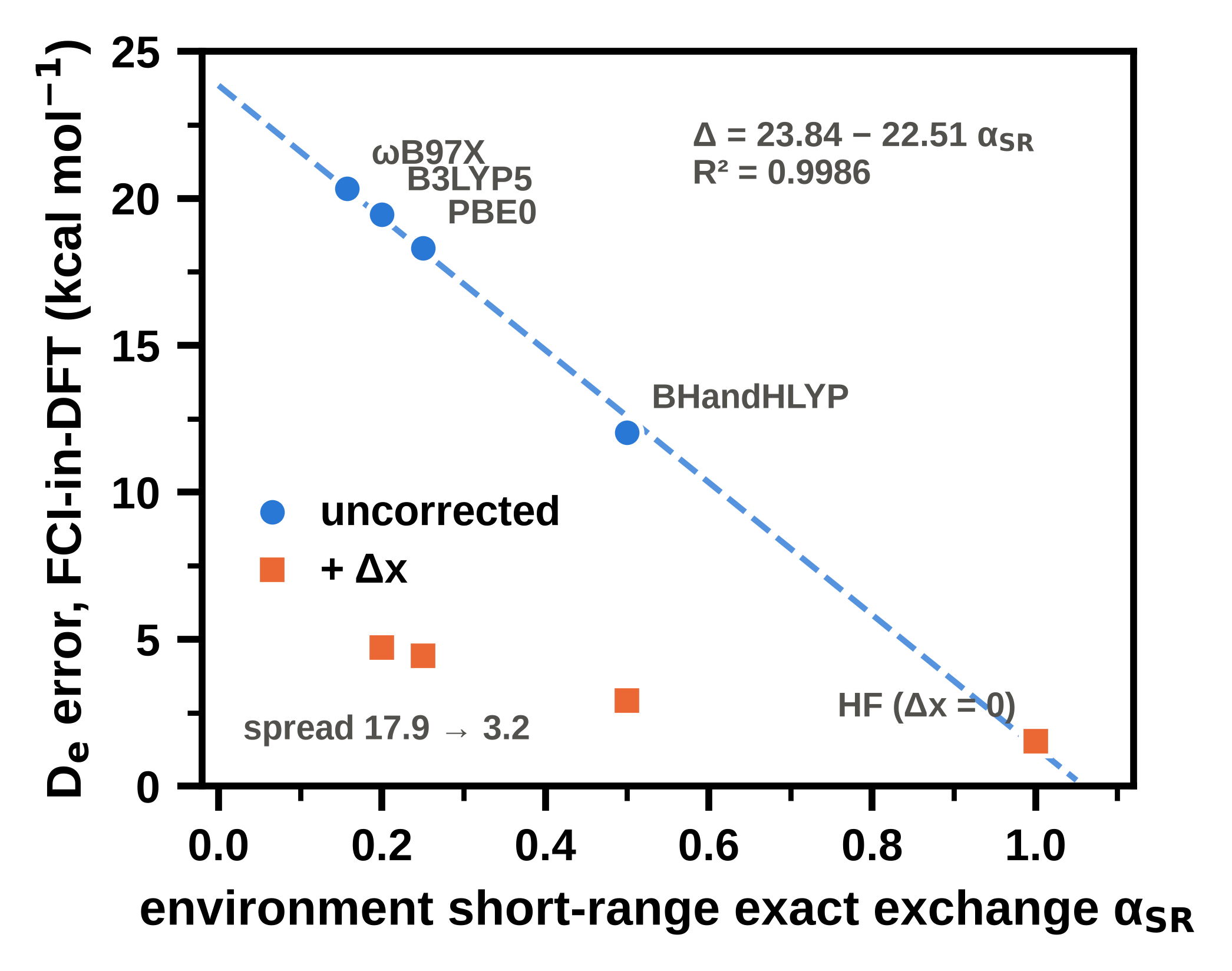}
\caption{Water embedding error (FCI-in-DFT, STO-3G) against the short-range exact-exchange fraction $\alpha_\mathrm{SR}$ of the environment functional, with restricted references. The uncorrected errors of five functionals (circles) are linear in $\alpha_\mathrm{SR}$ (dashed line, $R^2$ = 0.9986). With the nonadditive exchange correction $\Delta x$ (squares; four functionals, no $\Delta x$ point for $\omega$B97X), the spread falls from 17.9 to 3.2\,\kcal{}. For HF the correction vanishes.}
\label{fig:water}
\end{figure}

We use the dissociation-energy error $\Delta = D_e(\text{FCI}) - D_e(\text{FCI-in-DFT})$ along one O--H bond, with $D_e(\text{FCI})$ = 80.91\,\kcal{} (STO-3G). With restricted references, the error depends linearly on the short-range exact-exchange fraction $\alpha_\mathrm{SR}$ of the environment functional across five functionals spanning three correlation functionals (LYP, PBE and B97-type), $\Delta = 23.84 - 22.51\,\alpha_\mathrm{SR}$\,\kcal{} ($R^2$ = 0.9986; HF $+1.54$, BHandHLYP $+12.03$, PBE0 $+18.31$, B3LYP5 $+19.44$, $\omega$B97X $+20.34$; Figure~\ref{fig:water}). Semilocal functionals have no stable restricted solution on the stretched bond; with broken-symmetry references they extend the trend (PBE $+28.97$, SCAN $+25.46$), while the hybrids shift by only $+0.3$ to $+2.8$\,\kcal{} relative to their restricted values.

The partition is a larger error source than the functional. Moving a single occupied orbital between the subsystems changes $\Delta$ by $\sim$60\,\kcal{} (with SPADE forced to $n_\mathrm{act}$ = 5, 4, 3 and 2 active occupied orbitals, $\Delta$ = 0.0, $+19.4$, $-38$ and $-95$\,\kcal{} for B3LYP5), three times the entire functional range, and a partition that changes along the curve (IBO at a 0.99 threshold) gives meaningless dissociation energies. The partition is therefore held fixed and reported in every comparison below. With an exact (DMRG) embedded solver the B3LYP5 error is 19.44, 18.57 and 12.77\,\kcal{} in STO-3G, 6-31G and cc-pVDZ (PBE0: 18.31, 16.77, 11.81), while the exact $D_e$ rises from 80.9 to 102.1\,\kcal{}; the functional gap stays similar (1.1, 1.8 and 1.0\,\kcal{}), so the minimal basis does not hide the functional sensitivity, but the STO-3G partition numbers should be read as qualitative. CCSD is not a reliable embedded solver for this test: on the same embedded Hamiltonian it lies 13--14\,\mEh{} above FCI at 3--5\,\AA{} (6-31G), and at cc-pVDZ CCSD-in-DFT misplaces the error by $\sim$14\,\kcal{} by cancelling it against CCSD's own failure on the stretched bond.

The nonadditive exchange correction of eq~\ref{eq:dx} collapses the restricted-reference spread from 17.9 to 3.2\,\kcal{} (a factor of 5.6; BHandHLYP 2.92, PBE0 4.43, B3LYP5 4.72, HF unchanged at 1.54), so the functional dependence of the embedding error is almost entirely nonadditive exchange. The correction lowers the energy more at equilibrium than at dissociation ($-51$ vs $-28$\,\mEh{}, B3LYP5) and so increases $D_e$, correcting water in the right direction although the error has the opposite sign to the propionitrile case.\cite{monino2026jctc} For broken-symmetry references it is markedly less effective (spread 27.1 $\to$ 16.8\,\kcal{}, a factor of 1.6), indicating a substantial nonadditive correlation contribution. Iterating the embedding potential with the correlated active density (the SCF-DMRG-in-DFT variant, eq~19 of ref~\citenum{monino2026jctc}) converges in three macro-iterations for $r \leq 2$\,\AA{} but is ill-posed on the dissociating bond: at $r \geq 3$\,\AA{} the energy settles to the \uEh{} level while the density residual remains 0.4--0.9 under damping and DIIS, because the statically correlated active density switches between very different configurations at negligible energy cost. We therefore report one-shot energies.

\subsection{Periodic reference and cluster model}\label{sec:cluster}

\paragraph{Periodic holes.} The lowest cells are cfg04 at $x$ = 0.5 and cfg13 and cfg07 at $x$ = 1; the two $x$ = 1 cells lie within 5\,meV per cell in PBE+U. In every cell the extraction holes are on oxygen. The summed Bader spin of O equals the number of holes (1.94--2.12 for two holes, 3.79--3.99 for four), and Mn keeps a d$^3$ moment of $\approx$3\,$\mu_\mathrm{B}$. The Bader charge assigns only 50--65\,\% of the removed charge to O: about 0.75\,e is redistributed onto the Mn basins without any change of the Mn spin, increased Mn--O covalency rather than Mn oxidation. Each hole is shared by a symmetry-related O pair. The largest single-O spin is 0.33--0.47 at $x$ = 0.5 and 0.49--0.61 at $x$ = 1, so an active region holding a single oxygen would cut the hole in half by construction. HSE06 single points at the PBE+U geometries keep the energy ordering of the $x$ = 0.5 cells and the dominant hole-bearing pair in every cell, but localize the hole further, from 0.28--0.46 to 0.47--0.74 per O (site-projected moments, LORBIT = 11, projected onto the PAW projector functions). The functional thus changes \emph{how localized} the hole is, not \emph{where} it sits, and this is the quantity the embedding study probes.

\paragraph{Cluster convergence.} The cluster model was validated on cfg04 ($x$ = 0.5) against the periodic PBE+U hole at the DFT level (Table~\ref{tab:clusterconv}). At every size the holes sit on the same O sublattice as in the periodic calculation, and the next-largest spin on any other oxygen is 0.36. The spin on the target oxygen changes by only 0.016 between 3.6 and 4.5\,\AA{}. The cluster hole (0.57--0.58, L\"owdin/Mulliken) is more localized than the periodic PBE+U hole (0.39 site-projected, 0.47 Bader) and close to periodic HSE06 (0.53), although the different population measures make this ordering qualitative. Boundary oxygens, whose Mn partners are point charges, are more ionic than interior ones ($-1.1$ vs $-0.5$, Mulliken), but none approaches $-2$. The production calculations place no effective core potentials (ECPs) on the boundary cations, a common safeguard against electron leakage toward positive point charges.\cite{shi2022} In a test with Stuttgart small-core ECPs on the Mn and Li point charges within 4\,\AA{} of the QM region, the correlated R2 pair spins of cfg07 (HSE06, lowest state) change by 0.002\,e, and those of the 2.3\,\AA{} cluster by $\leq$0.022\,e (embedded R1--R3; $\leq$0.017\,e for the full-cluster DMRG). A small-core ECP does not represent the Mn 3s and 3p shells, so this test bounds only part of a possible leakage effect.
Because the hole-bearing sublattice reaches the cluster boundary, the production calculations use the 3.6\,\AA{} cluster, which keeps the active regions and their first neighbors interior, and the 4.5\,\AA{} cluster as a size check. For each active region the embedding is exact at the DFT-in-DFT level on the full cluster with point charges, DFT+U, density fitting and the broken-symmetry reference: the difference from the global energy is $\leq$0.03\,\uEh{} and $\langle S^2 \rangle$ is identical (Section~S3).

\begin{table}[tbp]
\caption{Hole Character of the cfg04 ($x$ = 0.5) Cluster against Cluster Size (PBE+U, DFT Level)}
\label{tab:clusterconv}
\centering
\begin{tabular}{cccccc}
\toprule
$r_\mathrm{cut}$ (\AA{}) & QM atoms & holes & spin on target O & hole-bearing O & Mn moments ($\mu_\mathrm{B}$) \\
\midrule
2.3 & 28 & 2 & 0.664 & O12/O13 images only & AFM, 3.2--3.3 \\
3.6 & 54 & 4 & 0.565 & O12/O13 images only & AFM, 3.1--3.4 \\
4.5 & 87 & 6 & 0.581 & O12/O13 images only & AFM, 3.1--3.3 \\
\bottomrule
\end{tabular}
\end{table}

\paragraph{Geometry and choice of the $x$ = 1 cell.} Hybrid and DFT+U functionals are known to disagree about O--O pairing in Li-rich oxides,\cite{chen2016,seo2016} so both $x$ = 1 cells were relaxed with HSE06 (Section~\ref{sec:periodic}). HSE06 forms no O--O dimer. In cfg13 the shortest O--O distance changes from 2.48 to 2.46\,\AA{} and the holes stay on the same four oxygens. In cfg07 the target O pair contracts from 2.581 to 2.409\,\AA{} (largest atomic displacement 0.40\,\AA{}), still far from a peroxo bond ($\sim$1.3--1.5\,\AA{}). What HSE06 does form is a hole polaron: in cfg13 the Mn--O bonds to the hole-bearing oxygens lengthen from 1.89--1.93 to $\approx$1.98\,\AA{}, those to spin-free oxygens shorten to 1.77\,\AA{}, and the hole moment rises from $\approx$0.45 to 0.76. The relaxation lowers the energy by 0.77\,eV (cfg07) and 0.78\,eV (cfg13), so the order of the two cells is unchanged: cfg07 lies 0.13\,eV per cell below cfg13 at the PBE+U geometry and 0.125\,eV below it at the HSE06 geometries. Because PBE+U puts cfg13 lower by 5\,meV and HSE06 puts cfg07 lower by 0.13\,eV, both cells are carried: cfg07 is the production baseline, and cfg13 serves for the configuration and cluster-size checks.

\subsection{Location and localization of the hole at fixed geometry}\label{sec:location}

\paragraph{Comparable correlated numbers.} A functional-by-region comparison is meaningful only if every number refers to the same electronic state, the same partition rule and converged solver settings. The protocols of Sections~\ref{sec:rulea} and \ref{sec:corr} were fixed before the numbers they govern were computed, with one exception, the extended state search described below. State tracking matters: for HF at $x$ = 0.5 the lowest candidate was internally unstable, and following the instability led to a stable solution of the same class 28.0\,\mEh{} below HF's periodic state, where discarding it would have reported a spurious change of state. The solver settings are converged for the hole spin. In R1 and R2 (92 runs) every bond-dimension error bar is $\leq$0.003\,e ($\leq$0.001\,e for the runs on the lower states found by the extended search), and a capped schedule agrees with direct large-$M$ runs to $\leq$0.0014\,e (23 and 34 orbitals). In R3, 13 of 22 runs have bars above 0.003\,e, most below 0.02\,e, with two reaching 0.068 (r2SCAN, 49/48) and 0.086 (HSE06 at the HSE06 geometry, 49/48) (Table~S2). With matched spin counts, the hole character shifts by $\leq$0.01\,e (R1) and $\leq$0.03\,e (R2) between NO thresholds 0.002 and 0.001; a third threshold for $x$ = 0.5, HSE06, R2 at fixed counts (32/28; 23, 33 and 39 orbitals) gives pair spins of $-0.906$/$-0.934$, $-0.937$/$-0.947$ and $-0.903$/$-0.931$, a shift of $\leq$0.03\,e with no drift. For the HSE06 state a at cfg07 (R2, 33/28; 27, 34 and 44 orbitals at thresholds 0.002, 0.001 and 0.0005) the pair spins are $-1.011$/$-0.999$, $-1.015$/$-0.995$ and $-0.994$/$-0.991$, a shift of $\leq$0.021\,e, with Mayer bond order $\leq$0.007 and O-2p share 0.50--0.51 throughout.
Energies, in contrast, do not converge comparably: extrapolations from the capped schedule and from large-$M$ points differ by 0.64\,\mEh{} (23 orbitals) and 5.8\,\mEh{} (34 orbitals).

\paragraph{Extended state search.} The initial search missed two stable states (Table~S10 and Section~S8 of the Supporting Information). Both occur at cfg07 in the PBE+U geometry, and both are lower than the initially recorded state a$'$ while keeping its Mn pattern and spin purity. For HSE06 the ECP anneal finds a state 2.24\,\mEh{} lower ($\langle S^2 \rangle$ 13.69 $\to$ 13.71); for BHandHLYP only the cross-seed from this HSE06 state finds a state 8.04\,\mEh{} lower ($\langle S^2 \rangle$ 15.39 $\to$ 15.47), whereas BHandHLYP's own anneal returns to state a$'$. In both, the pair hole becomes nearly full and uncoupled: the SCF pair-O spins (IAO) change from $-0.75$/$-0.90$ to $-0.94$/$-0.98$ (HSE06) and from $-0.60$/$-1.05$ to $-0.97$/$-1.06$ (BHandHLYP). At the correlated level the change reaches 0.20\,e (HSE06, R1) and 0.45\,e (BHandHLYP, R1) on one oxygen, and the O--O covalency of state a$'$ disappears (Mayer bond order 0.07 $\to$ $\leq$0.005 for HSE06). Every other case returns to, or lands above, the recorded state a: PBE+U at all three $U$ values, r2SCAN (its HSE06 seed lands 0.59\,\mEh{} higher) and HF at cfg07 (anneal and cross-seed), HSE06 at cfg13, HSE06 and BHandHLYP at the HSE06 geometry, and the periodic-pattern hybrid states at $x$ = 0.5; the one exception, for HSE06 at the HSE06 geometry, is a solution 0.28\,\mEh{} lower that coincides with the recorded state in R1--R3 (Section~\ref{sec:geometry}). A periodic HSE06 calculation of cfg07 seeded with fully localized pair holes returns to the published periodic state, but the four-formula-unit cell cannot represent the symmetry-broken hole arrangement of the new cluster state, so this test is inconclusive. Following the protocol, the lower states are state a from here on; state a$'$ is retained in the Supporting Information and as the reference state of the SQD tests (Section~\ref{sec:sqd}). Neither route alone finds both states, so the search needs both the anneal and cross-seeding; the missed states are also an entry of the error budget (Section~\ref{sec:budget}).

\begin{figure}[tbp]
\centering
\includegraphics[width=\textwidth]{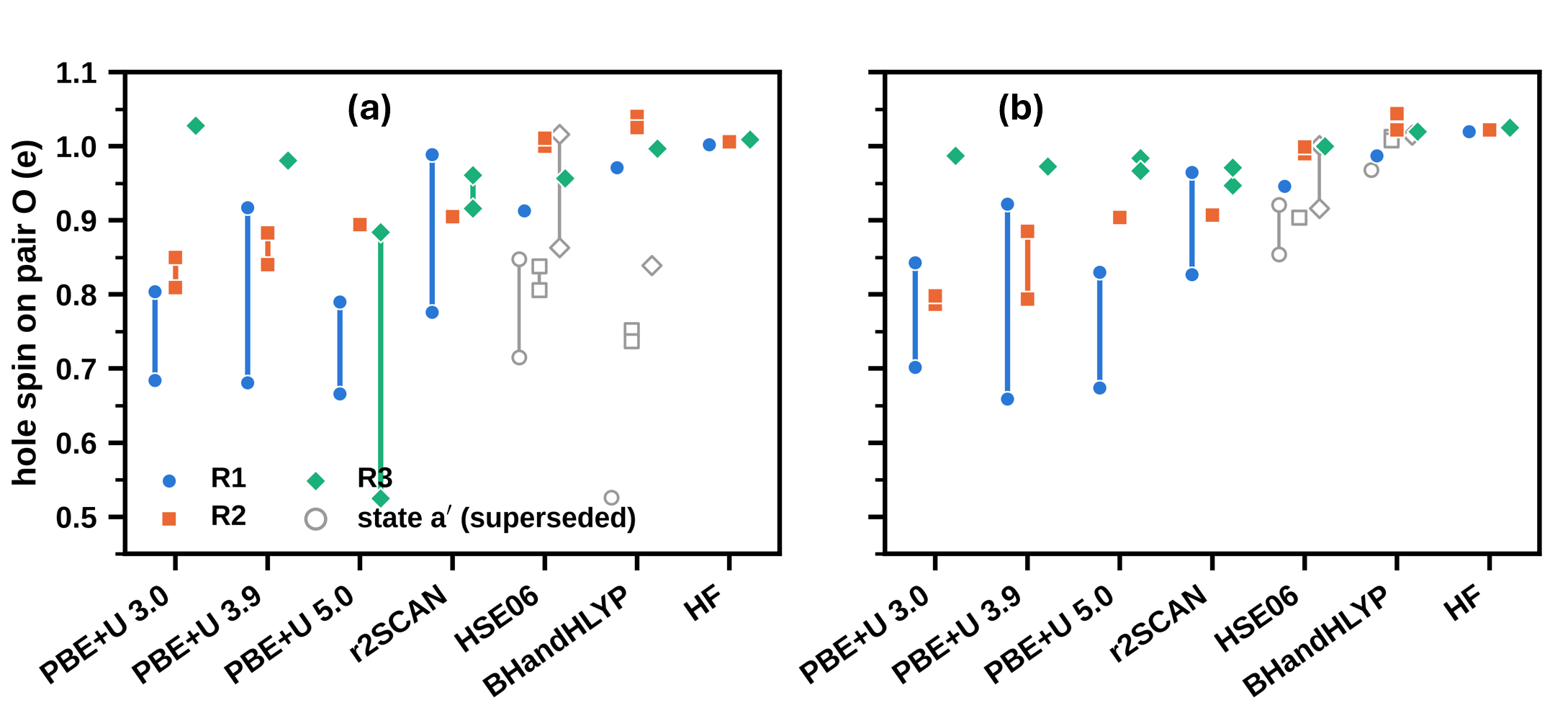}
\caption{Hole spin (negative of the pair-O spin) on the two oxygens of the target pair, (a) and (b), from DMRG-in-DFT ($x$ = 1, cfg07, PBE+U geometry, NO threshold 0.002), for seven environment functionals and the three nested active regions (R1, the O pair; R2, R1 and its two Mn; R3, R2 and the neighboring hole-bearing and bridging O). Filled symbols, state a (for HSE06 and BHandHLYP the lower states of the extended search); open gray symbols, the initially recorded states a$'$ of HSE06 and BHandHLYP (one BHandHLYP R1 option of state a$'$, pair-O spin $+0.98$, lies off scale). Where the spin inside the active region is not an integer, both admissible active-space spin projections are computed (rule A, Section~\ref{sec:rulea}) and the bar spans the two results. Bond-dimension errors are smaller than the symbols for R1 and R2 and reach 0.07--0.09\,e for two R3 runs (Table~S2).}
\label{fig:holegrid}
\end{figure}

\paragraph{The hole sits on the O pair.} At the PBE+U geometry, the lowest stable state of all seven functionals (PBE+U with $U$ = 3.0, 3.9 and 5.0\,eV, r2SCAN,\cite{r2scan} HSE06,\cite{hse06} BHandHLYP\cite{bhandhlyp} and HF) has the hole on the O pair, at $x$ = 0.5 and $x$ = 1 (Figure~\ref{fig:holegrid}). How strongly the hole localizes depends on the functional, and at this fixed geometry the functionals fall into two groups. At $x$ = 1 in R2 the pair-O spin magnitudes, over all spin-count options and both oxygens, are 0.79--0.90 for PBE+U (rising with $U$) and 0.91 for r2SCAN, but 0.99--1.01 for HSE06, 1.02--1.04 for BHandHLYP and 1.01--1.02 for HF. DFT+U and the semilocal functional give a partial hole; every functional with exact exchange gives a full one. The O--O covalency follows the same split: the R2 Mayer bond order is 0.06--0.12 for PBE+U, 0.014 for r2SCAN and $\leq$0.005 for the hybrids and HF. The geometry, however, changes this picture for the hybrids (Section~\ref{sec:geometry}).

\paragraph{At $x$ = 0.5 the cluster prefers a Mn pattern that the crystal does not.} For every functional (HF after instability following), the lowest cluster state at $x$ = 0.5 is the O pair with a non-periodic Mn pattern and reversed hole-spin orientation, in which two periodic images of the same Mn carry opposite spins. Its preference over the periodic-pattern state grows with exact exchange, from 0.5\,\mEh{} for PBE+U ($U$ = 3.0\,eV) to 28.0\,\mEh{} for HF (Table~\ref{tab:x05}). We tested this pattern in the crystal with periodic PBE+U ($U$ = 3.9\,eV) in a $2\times1\times1$ supercell at the same geometry, the smallest cell that can hold it. Started from the cluster pattern, with either sign of the oxygen moments, the calculation keeps the flipped Mn and the reversed hole spins, so the state exists, but it lies 0.26\,eV (9.6\,\mEh{}) above the periodic pattern, where the cluster placed it 1.2\,\mEh{} below. The cluster preference is therefore an artifact of the finite cluster, and at $x$ = 0.5 the periodic-pattern state is the physical one. In that state the semilocal and DFT+U functionals give a nearly symmetric, nearly full hole, while the hybrids make the pair asymmetric (Table~\ref{tab:x05}); the extended search confirms the asymmetric hybrid states (a symmetric HSE06 solution lies 0.76\,\mEh{} higher). The hole stays on the O pair in both Mn patterns, but its character depends on the Mn pattern more than on the functional, by up to 0.7\,e on one oxygen for HSE06 (0.91--0.93 on both oxygens in the cluster-lowest pattern). Hole character at $x$ = 0.5 can thus be compared only within one magnetic state, and that state must be checked against the periodic crystal.

\begin{table}[tbp]
\caption{The $x$ = 0.5 Cluster: Preference for the Non-Periodic Mn Pattern and Pair-O Spin in the Periodic-Pattern State (R2)}
\label{tab:x05}
\centering
\begin{tabular}{lcc}
\toprule
functional & $E_\text{periodic} - E_\text{cluster-lowest}$ (\mEh{}) & $|$pair-O spin$|$, periodic pattern \\
\midrule
PBE+U, $U$ = 3.0 & 0.5 & \multirow{3}{*}{0.90--1.00} \\
PBE+U, $U$ = 3.9 & 1.2 & \\
PBE+U, $U$ = 5.0 & 2.7 & \\
r2SCAN & 2.8 & 0.96--1.12 \\
HSE06 & 5.3 & 0.26 / 0.93 \\
BHandHLYP & 11.0 & 0.60 / 0.93 \\
HF & 28.0 & 0.95 / 1.03\textsuperscript{\emph{a}} \\
\bottomrule
\end{tabular}

\raggedright\footnotesize\textsuperscript{\emph{a}}SPADE orbital counts.
\end{table}

\subsection{The active-region boundary and the active-space spin}\label{sec:boundary}

\begin{figure}[tbp]
\centering
\figorbox{\textwidth}{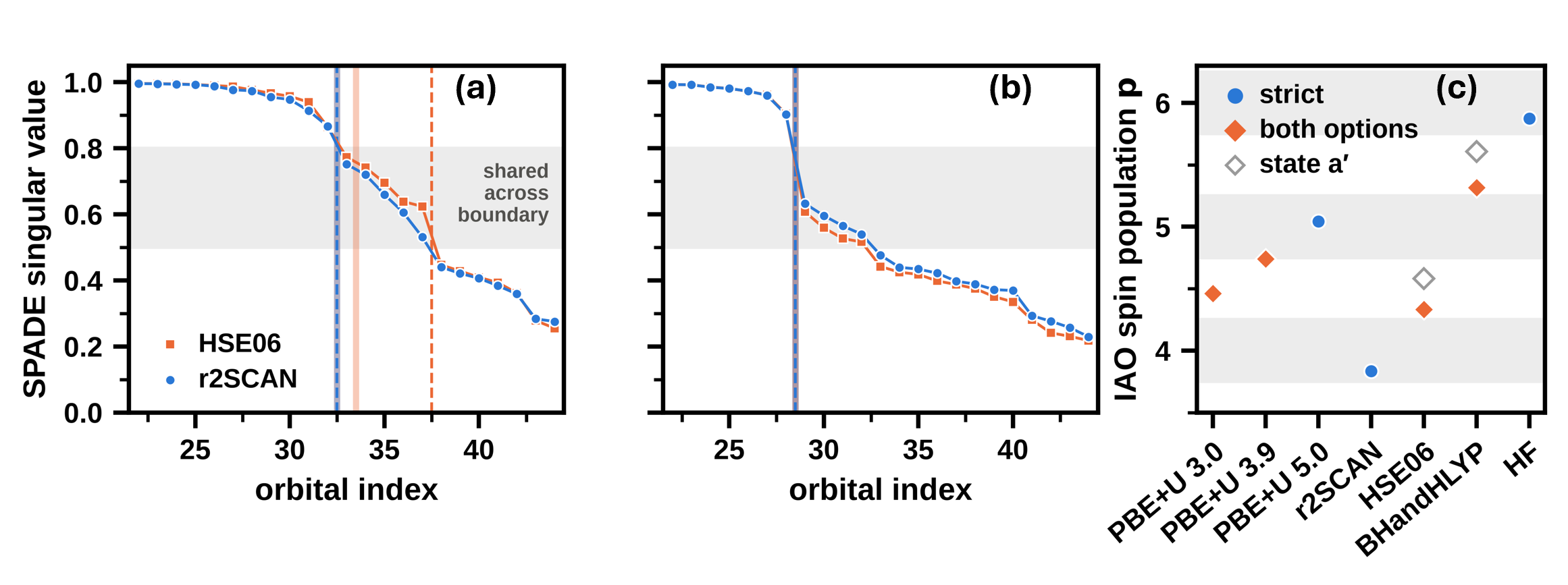}{5.5cm}
\caption{SPADE singular values of the R2 partition ($x$ = 1, cfg07, PBE+U geometry) for the (a) $\alpha$ and (b) $\beta$ occupied spaces with HSE06 (state a$'$) and r2SCAN environments, shown around the cut, with the orbitals in order of descending singular value. Dashed lines mark SPADE's largest-gap cut and solid lines the cuts selected by rule A; the shaded band marks singular values of 0.5--0.8, where orbitals are shared across the boundary and the largest-gap criterion becomes ambiguous. (c) IAO spin population $p$ of the R2 atoms (eq~\ref{eq:p}) in state a for the seven environment functionals (for HSE06 and BHandHLYP the lower states of the extended search; open gray diamonds, their states a$'$); the shaded bands are the $\pm$0.25 intervals around an integer within which rule A fixes a single \Ms{} (circles), and outside them both bracketing values are computed (diamonds).
}
\label{fig:spade}
\end{figure}

\paragraph{The functional moves spin across the boundary.} The ambiguity that motivates rule A is visible in the SPADE spectra themselves (Figure~\ref{fig:spade}): with nearly identical spectra, HSE06 (state a$'$) and r2SCAN cut $\alpha$ at 37 and 32 orbitals in R2 ($2M_S$ = +9 vs +4). Once the counts are tied to the SCF state, the matched counts within one state class and region differ between functionals by about one $\alpha$ orbital, following the SCF spin inside the region, which grows with exact exchange ($x$ = 1, R2: r2SCAN 3.8, HSE06 4.3, BHandHLYP 5.3, HF 5.9; 4.6 and 5.6 in the states a$'$ of HSE06 and BHandHLYP). The environment functional thus moves spin across the active/environment boundary.

\paragraph{The spin-count choice matters mainly when the active region holds only the oxygen pair.} The two rule-A options differ most in R1, where the region is the O pair alone and its Mn--O bonds are cut. At the PBE+U geometry the largest difference on a pair oxygen is 0.14, 0.26 and 0.16\,e for PBE+U at $U$ = 3.0, 3.9 and 5.0\,eV and 0.21\,e for r2SCAN. Once the two Mn are included (R2), it drops to 0.04 and 0.09\,e for PBE+U at $U$ = 3.0 and 3.9\,eV (single admissible count for $U$ = 5.0\,eV and r2SCAN) and to 0.011 and 0.023\,e for HSE06 and BHandHLYP, whose R1 and R3 have a single admissible count. In R3 it is 0.05\,e for r2SCAN and 0.36\,e for PBE+U at $U$ = 5.0\,eV; there $p$ = 2.29 lies only 0.04 outside the strict interval, and the far option ($2M_S$ = 3, 0.71 from $p$) moves one oxygen to $-0.53$. HF has a single admissible count in every region. The hybrids reach this clean regime only at the PBE+U geometry: at the HSE06 geometry HSE06 is a spread case in all three regions (differences 0.08, 0.024 and 0.23\,e in R1, R2 and R3), and for BHandHLYP the R1 choice decides the coupling itself, giving a parallel pair ($-0.29$/$-0.92$, 8/10) or an antiparallel one ($+1.07$/$-0.92$, 9/10, SPADE's own cut); both are admissible ($p$ = $-1.36$), so the integer \Ms{} of the active space, not the environment, decides the coupling. Adding the two Mn neighbors of the pair (R2) therefore reduces the spin-count choice to $\leq$0.09\,e for PBE+U and HSE06 at their own geometries; in the larger R3 it can again reach 0.23\,e (HSE06 at its own geometry) and 0.36\,e (PBE+U, $U$ = 5.0\,eV).

\paragraph{Boundary steps.} Each step R1 $\to$ R2 $\to$ R3 is reported as an interval per pair oxygen that spans every spin-count option of both regions and is widened by the truncation and bond-dimension bars (Figure~\ref{fig:holegrid}, Table~S3). For PBE+U the resolved steps, which exclude zero, are R2 $\to$ R3 at $U$ = 3.0\,eV (both oxygens, $-0.27\ldots-0.13$\,e) and 3.9\,eV (both oxygens, $-0.24\ldots-0.03$), and R1 $\to$ R2 (both oxygens, $-0.25\ldots-0.06$) and R2 $\to$ R3 (one oxygen, $-0.13\ldots-0.02$) at $U$ = 5.0\,eV; all of r2SCAN's include zero. For the hybrids and HF in state a every step is $\leq$0.10\,e (HSE06: R1 $\to$ R2 $-0.10\ldots-0.04$, R2 $\to$ R3 $-0.01\ldots+0.05$; BHandHLYP $\leq$0.07; HF within $\pm$0.05). In state a$'$ the HSE06 R2 $\to$ R3 interval reached $-0.27$\,e because one R3 option (50/48) already gave a full hole ($-1.016$/$-1.001$): the larger correlated region recovered part of the localization that the state a$'$ SCF lacked. For PBE+U, where the anneal and the cross-seed confirm state a as the lowest SCF state, the large R2 $\to$ R3 steps are already present in the embedded mean field ($-0.18\ldots-0.21$\,e at $U$ = 3.0\,eV, against $-0.18\ldots-0.22$\,e with DMRG; Section~\ref{sec:levels}): they reflect the exact exchange that the embedding applies inside the larger active region, which favors a more localized hole than the DFT+U environment, rather than correlation. Neither the boundary nor the spin assignment has converged by R3 for DFT+U.

\paragraph{Embedded vs full-cluster DMRG.} On the small 2.3\,\AA{} cluster (22 atoms) a full-cluster DMRG is feasible. The valence space of all atoms (Mn 3d and O 2p, 40 orbitals per spin, converged at $M$ = 1000) gives pair spins of $+0.882$/$+0.943$ (the sign is the hole-spin orientation of this cluster's state; Table~S6). With the embedded runs restricted to the valence space of their own region (AVAS\cite{avas2017}), R1 gives $+0.861$/$+0.911$, 0.02--0.03\,e below the full cluster, and R2 gives $+0.888\ldots+0.895$/$+0.946\ldots+0.950$, within 0.013\,e; with the production MP2 natural-orbital spaces the deviations are 0.03--0.04\,e (R1) and $\leq$0.013\,e (R2). R3 overshoots in both spaces, by 0.14 and 0.06\,e on the two oxygens (valence space $+1.025$/$+1.006$), so the overshoot is not an active-space effect. On this small cluster R3 reaches the boundary: four of its eight atoms, including one pair oxygen, are bonded (1.9\,\AA{}) to bare $+4$ point charges. With small-core ECPs on the boundary cations the overshoot persists (0.15/0.08\,e), so as far as this test can show it is not caused by electron leakage toward the bare charges; it more likely reflects the truncation of R3 at the cluster edge. In the 3.6\,\AA{} production cluster only one R3 oxygen, not a pair oxygen, is in such contact, and the pair oxygens are $\geq$3.1\,\AA{} from the nearest point charge. We therefore regard R2 as the region with the cleanest boundary.

\subsection{The hole character depends on the functional together with the geometry}\label{sec:geometry}

At the HSE06-relaxed geometry the R2 hole description first appeared to change by up to $\sim$0.7\,e, with three functionals changing sign or spin pattern. Two artifacts produced most of this (Section~S6, P1 and P2, of the Supporting Information): at a fixed $r_\mathrm{cut}$ the relaxed cell drops one O image from the cluster (44 $\to$ 43 atoms, charge $+1 \to +3$), and the PBE+U-geometry seeds failed on the changed cluster, so that each SCF found one unseeded solution 43--300\,\mEh{} above the lowest state. We therefore carve the HSE06-geometry cluster from the same atoms (site and periodic image) as the PBE+U one and repeat the full state search, including the extended search for HSE06 and BHandHLYP. At the matched geometry the lowest stable state of all seven functionals is again an O-pair state with the periodic Mn pattern. For HSE06 the HF- and BHandHLYP-seeded searches converge to a solution 0.28\,\mEh{} below the recorded state; its IAO spin populations $p$ (R1 $-1.553$, R2 4.466, R3 1.666 against $-1.552$, 4.467 and 1.660), spin counts and atomic spins inside R1--R3 ($\leq$0.004\,e) coincide with the recorded state, and it differs only on environment oxygens 3.3--4.8\,\AA{} from the pair ($\leq$0.014\,e), so the existing DMRG results stand. The fully localized HSE06 state lies 9.14\,\mEh{} higher at this geometry, and for BHandHLYP the localized state lies 3.85\,\mEh{} higher.

\begin{figure}[tbp]
\centering
\includegraphics[width=\textwidth]{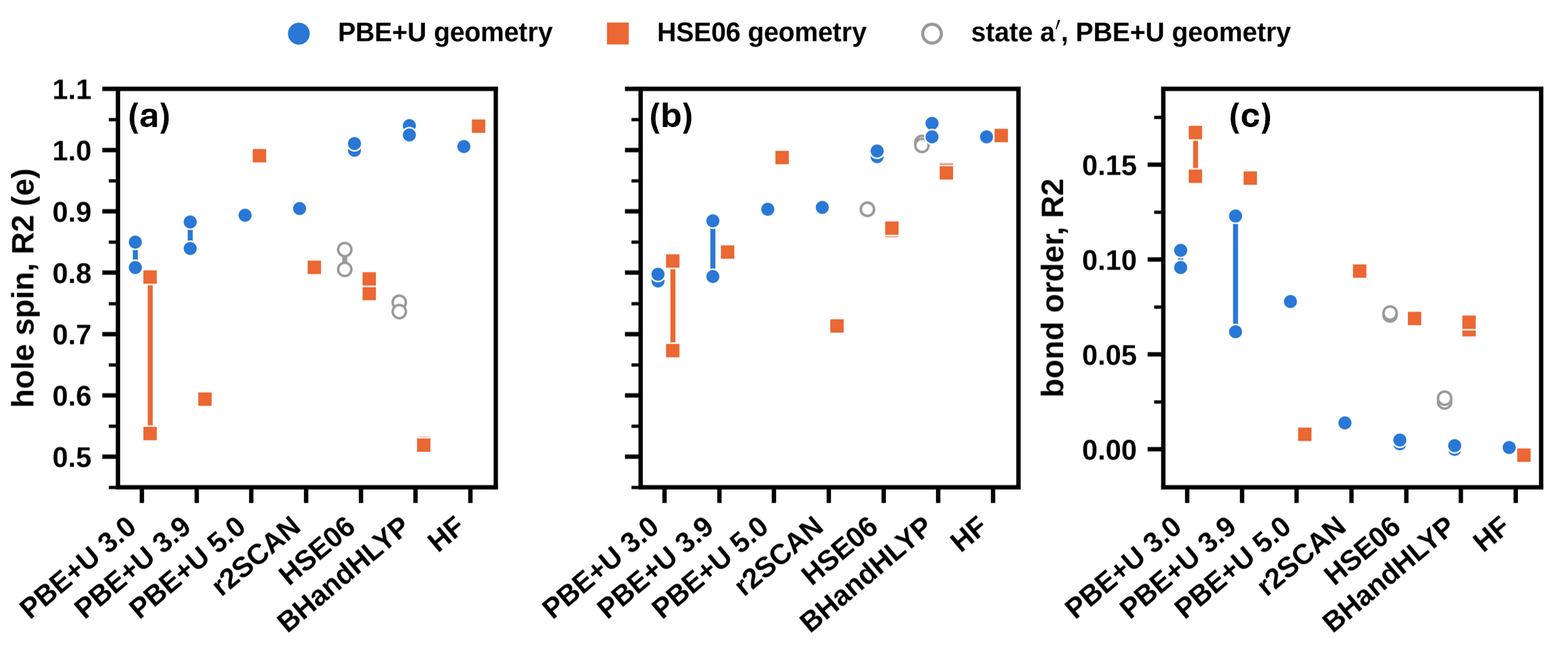}
\caption{Effect of the geometry on the R2 hole ($x$ = 1, cfg07, NO threshold 0.002), with the clusters at the PBE+U and the HSE06 geometry carved from the same atoms. (a, b) Hole spin (negative of the pair-O spin) on the two pair oxygens and (c) O--O Mayer bond order, for the lowest stable state of each environment functional at the PBE+U geometry (circles) and at the HSE06 geometry (squares); bars span the two rule-A spin-count options. Open gray circles, the initially recorded states a$'$ of HSE06 and BHandHLYP at the PBE+U geometry. HF is insensitive to the geometry; the hybrids change most, from a full, uncoupled hole at the PBE+U geometry to a partial (HSE06) or asymmetric (BHandHLYP), weakly covalent hole at the HSE06 geometry.}
\label{fig:geometry}
\end{figure}

The geometry then changes the hole by an amount that depends on the functional (Figure~\ref{fig:geometry}, Table~S4; R2, lowest states). HF is insensitive: $-1.01$/$-1.02$ at the PBE+U geometry and $-1.04$/$-1.02$ at the HSE06 geometry. PBE+U at $U$ = 5.0\,eV localizes further ($-0.89$/$-0.90$ $\to$ $-0.99$/$-0.99$), r2SCAN moves by 0.1--0.2\,e ($-0.905$/$-0.907$ $\to$ $-0.809$/$-0.713$), and with PBE+U at $U$ = 3.0 and 3.9\,eV the hole becomes unequal on the two oxygens, one of which moves by up to 0.3\,e in R2 and 0.4\,e in R3 ($U$ = 3.9\,eV: R2 $-0.88$/$-0.89$ $\to$ $-0.59$/$-0.83$; R3 $-0.98$/$-0.97$ $\to$ $-1.01$/$-0.55$). The hybrids change most. HSE06 goes from a full hole at the PBE+U geometry ($-1.00\ldots-1.01$/$-0.99\ldots-1.00$) to a partial one at its own geometry ($-0.77\ldots-0.79$/$-0.87$), a change of 0.21--0.25 and 0.12--0.13\,e on the two oxygens; BHandHLYP goes from $-1.03\ldots-1.04$/$-1.02\ldots-1.04$ to an asymmetric pair, $-0.52$/$-0.96\ldots-0.97$, a change of 0.50--0.52\,e on one oxygen, with $-0.29$/$-0.92$ in R1 (8/10) and $-0.61$/$-0.99$ in R3. A mirror-image BHandHLYP solution with the asymmetry on the other oxygen (SCF level $-0.94$/$-0.33$) lies 5.1\,\mEh{} higher, and further mirror-type solutions lie 5--12\,\mEh{} higher (Supporting Information). At both geometries BHandHLYP polarizes spin more strongly than HSE06 (Mn moments of 3.3--3.6 against 3.0--3.2\,$\mu_\mathrm{B}$, and about 2\,e more spin summed over the other oxygens).

The O--O covalency follows the geometry more than the functional (Figure~\ref{fig:geometry}c). At the PBE+U geometry the R2 Mayer bond order is $\leq$0.005 for the hybrids and HF, 0.014 for r2SCAN and 0.06--0.12 for PBE+U. At the HSE06 geometry, where the pair has contracted from 2.58 to 2.41\,\AA{}, it is 0.06--0.17 for every environment except HF and PBE+U at $U$ = 5.0\,eV ($\leq$0.01). For HSE06 at its own geometry the value is the same in R2 and R3 (0.069 and 0.06--0.07), whereas for PBE+U at its own geometry it falls from 0.06--0.12 in R2 to 0.00--0.05 in R3, and for BHandHLYP at the HSE06 geometry from 0.065 to 0.025. Weak O--O covalency of about 0.07 is therefore converged with region size only for the geometry-consistent hybrid.

Taken together, the full, uncoupled hole that the hybrids give at the PBE+U geometry is a property of the combination of a hybrid functional with the longer O--O distance of the DFT+U geometry. At their own, contracted geometry the hybrids return to a partial or asymmetric, weakly covalent hole whose spin spreads across the boundary, as PBE+U does at its own geometry. A functional evaluated at another functional's geometry can move the hole by up to 0.5\,e on one oxygen in R2.

\subsection{Energies, competing states and the magnetic reference}\label{sec:energies}

The configuration energy is not converged: $E(\text{cfg07}) - E(\text{cfg13})$ moves by 5--47\,\kcal{} between NO thresholds 0.002 and 0.001, and the second-order (MP2) correction for the natural-orbital truncation does not repair it. The hole character converges at $\sim$30--40 correlated orbitals; energies do not, so configuration energies are not reported as results.

The HSE06 SCF at the PBE+U geometry has a stable solution with the hole on a different oxygen, but none of the six other environments retains it. Started from the HSE06 other-O density, HF and BHandHLYP relax to O-pair solutions (with a different Mn pattern), 35.8\,\mEh{} above HF's state a and 11.4\,\mEh{} above BHandHLYP's state a$'$ (19.4\,\mEh{} above its state a), and r2SCAN and PBE+U at all three $U$ values do the same, 14--17\,\mEh{} above their state a. In no environment is the seeded state the lowest, so at the SCF level the O pair is the hole location for every functional. At the correlated level the ranking is not resolved (Table~S11; R2, matched spin counts 33/28 for both states at every threshold, MP2-corrected). Against the HSE06 state a, the other-O state lies $+2.66$, $-1.79$ and $-3.24$\,\kcal{} relative to the O pair at NO thresholds 0.002, 0.001 and 0.0005 with the nonadditive exchange correction ($\Delta x$, eq~\ref{eq:dx}), and $+10.07$, $+5.38$ and $+4.05$\,\kcal{} without it. Both series drift by about 6\,\kcal{} over the three thresholds, so the gap is not converged with the virtual space, and with $\Delta x$ it changes sign; without the MP2 correction the $\Delta x$-included gaps ($+10.76$, $-9.68$ and $-0.35$\,\kcal{}) do not even follow a trend. The $\Delta x$ contribution itself is stable along each series: $-7.41$, $-7.17$ and $-7.29$\,\kcal{} against state a, and $+34.35$, $+34.72$ and $+34.17$\,\kcal{} against state a$'$, against which the other-O state had lain 7.4--11.5\,\kcal{} above the O pair with $\Delta x$ and 22.9--26.7\,\kcal{} below it without. The $\Delta x$ contribution to the relative energy therefore changes by about 41\,\kcal{} between two O-pair states that differ by only 1.4\,\kcal{} at the SCF level. Because $\Delta x$ enters the energy only, it cannot move the hole within a state, but its contribution to the relative energy of two open-shell states depends strongly on which states, and which partitions, are compared. Rankings of competing open-shell states with and without $\Delta x$ at feasible active-space sizes should therefore be reported as unresolved unless both agree, which is not the case here.

With all Mn parallel (a stable high-spin reference of the same O-pair class, 46.9\,\mEh{} above the broken-symmetry state a$'$ of HSE06 at the SCF level), R2 gives $+0.73$/$-0.99$: the two pair oxygens are antiparallel, against $-0.81$/$-0.90$ (parallel) in the broken-symmetry state a$'$. The hole stays on the pair, but its spin pattern follows the Mn magnetic order, so the magnetic reference is a setting that must be reported.

\subsection{Error budget for the hole character}\label{sec:budget}

\begin{figure}[tbp]
\centering
\includegraphics[width=\textwidth]{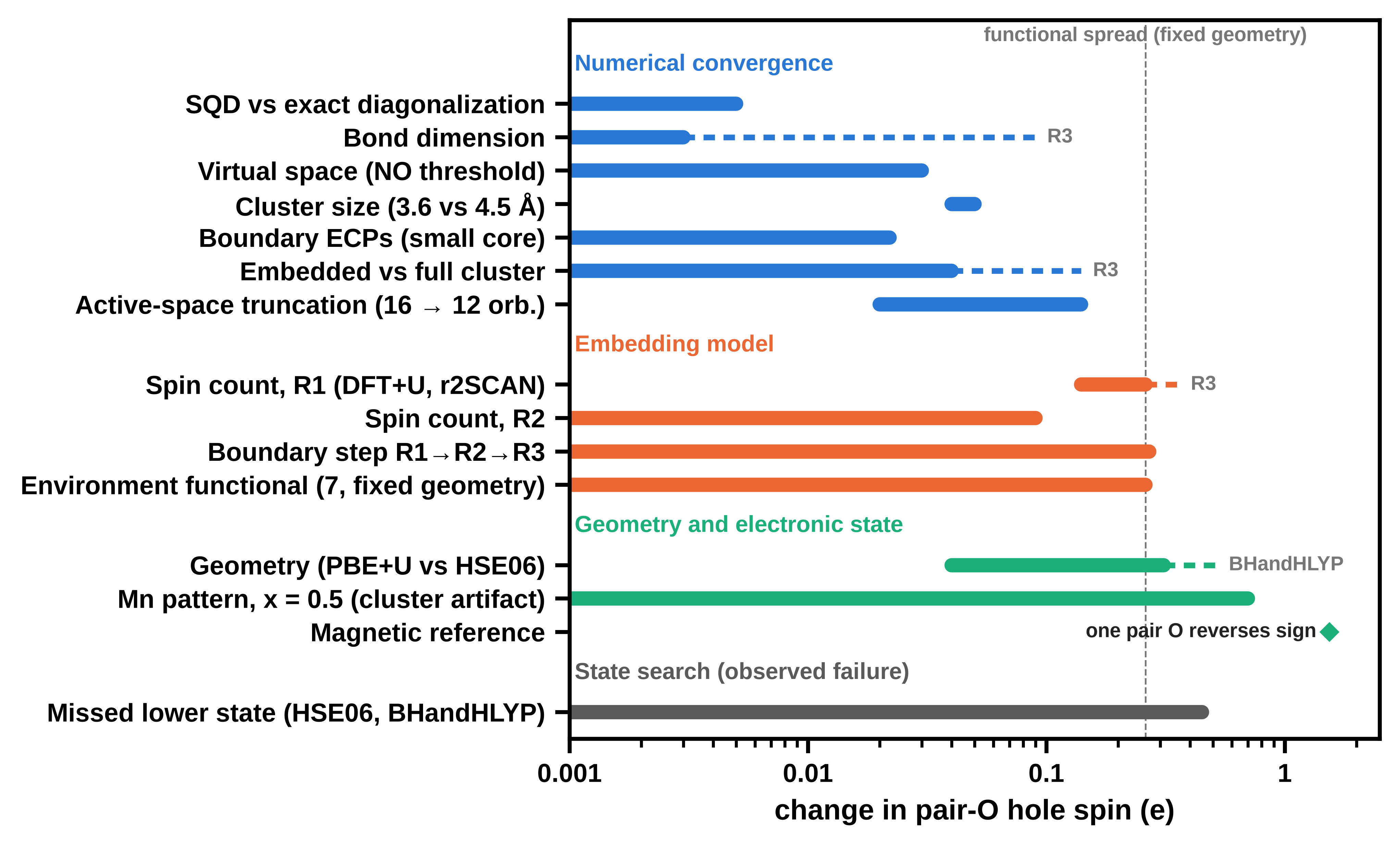}
\caption{Error budget for the pair-O hole spin at $x$ = 1 (Table~\ref{tab:budget}; logarithmic axis). Bars span the reported range of each source; bars that start at the left edge are upper bounds. Dotted extensions show larger values in R3 or, for the geometry, BHandHLYP. The dashed line marks the spread over the seven environment functionals at the PBE+U geometry. Numerical settings (blue) are converged well below the embedding-model choices (orange); geometry and electronic-state choices (green) and the observed failure of the initial state search (gray) exceed them.}
\label{fig:budget}
\end{figure}

Table~\ref{tab:budget} and Figure~\ref{fig:budget} collect the changes of the pair-O spin and of the composition measures, grouped by their origin. The numerical settings (bond dimension, virtual space, cluster size, boundary ECPs, solver) are converged to a few hundredths of an electron in R1 and R2. The embedding-model settings (functional at fixed geometry, boundary, active-space spin count) each move the pair-O spin by up to 0.26--0.27\,e. The geometry moves it by up to 0.31\,e for DFT+U and 0.5\,e for BHandHLYP, and the electronic-state choices (Mn pattern, magnetic reference) can move it further or reverse its sign. The two states missed by the initial search changed a pair-O spin by up to 0.45\,e; this entry records a failure of the search rather than a sensitivity to a setting, and it is shown separately. Two observations follow from the composition columns. First, no O--O bond forms in any setting: the Mayer bond order of the pair is $\leq$0.24 everywhere and $\leq$0.14 at the PBE+U geometry, against $\sim$1 for a peroxo bond. What the settings decide is weak O--O covalency, which is set mainly by the geometry (Section~\ref{sec:geometry}) and is the observable most sensitive to the treatment of the boundary: geometry, region and spin count each move it by about its own size. Second, the O-2p share is stable where the spin is stable: the NO threshold, bond dimension, cluster size and magnetic reference move it by $\leq$0.02, while the functional ($\sim$0.19) and the spin-count option ($\leq$0.19) move it as much as they move the spin. Every run additionally reports the spread between its two DMRG starts, $\langle S^2 \rangle$, the active-space size and per-spin counts with the outcome of the \Ms{} check, and the state label and its stability (Tables~S1--S3 and S10).

\begin{table}[tbp]
\caption{Error Budget for the Hole Character at $x$ = 1\textsuperscript{\emph{a}}}
\label{tab:budget}
\centering
\footnotesize
\begin{tabularx}{\textwidth}{>{\raggedright\arraybackslash}p{2.5cm} >{\raggedright\arraybackslash}p{2.3cm} >{\raggedright\arraybackslash}X >{\raggedright\arraybackslash}p{2.2cm} >{\raggedright\arraybackslash}p{2.5cm}}
\toprule
source & values & $\Delta$ pair-O spin (e) & $\Delta$ O-2p share & $\Delta$ O--O bond order \\
\midrule
\multicolumn{5}{l}{\emph{numerical convergence}} \\
NO threshold & 0.002 / 0.001 / 0.0005 & $\leq$0.01 (R1), $\leq$0.03 (R2; $\leq$0.021 for the HSE06 state a) & $\leq$0.01 (R2) & $\leq$0.002 (R1), $\leq$0.011 (R2) \\
bond dimension & $M$ schedule & $\leq$0.003 (R1, R2); $\leq$0.09 (R3) & $\leq$0.008 & $\leq$0.009 \\
cluster size & 3.6 / 4.5\,\AA{} (cfg13, R2) & 0.04--0.05 (same Mn pattern) & $\leq$0.001 & $\leq$0.007 \\
boundary ECPs & small core, within 4\,\AA{} & 0.002 (R2); $\leq$0.022 (2.3\,\AA{} cluster; full cluster $\leq$0.017) & -- & -- \\
embedded vs full cluster & 2.3\,\AA{}, valence space & $\leq$0.04 (R1), $\leq$0.013 (R2); R3 overshoots by $\leq$0.14 & -- & -- \\
solver and active space & DMRG (25 NOs) / 12--16-orbital reduced active space, SQD & truncation 0.13--0.14 (12) $\to$ 0.02--0.06 (16); SQD vs FCI $\leq$0.005 & -- & -- \\
\midrule
\multicolumn{5}{l}{\emph{embedding model (PBE+U geometry unless noted)}} \\
functional & 7 & 0.79--1.04 ($\sim$0.26) & 0.48--0.67 ($\sim$0.19) & $\leq$0.005 (hybrids, HF) -- 0.12 (PBE+U) \\
boundary step & R1 / R2 / R3 & $\leq$0.27 (PBE+U); $\leq$0.10 (hybrids, HF) & R2 $\to$ R3: $-0.14\ldots+0.05$ & $-0.10\ldots+0.13$ \\
active-space spin count & both rule-A options & R1: 0.14--0.26 (PBE+U, r2SCAN); R2: $\leq$0.09; R3: $\leq$0.36\textsuperscript{\emph{b}}; hybrids $\leq$0.02; HSE06 at its geometry 0.08 / 0.02 / 0.23 & median 0.02, $\leq$0.19 & median 0.004, $\leq$0.08 (R2/R3); $\leq$0.18 (R1) \\
\midrule
\multicolumn{5}{l}{\emph{geometry and electronic state}} \\
geometry & PBE+U / HSE06 & $\leq$0.04 (HF); 0.08--0.31 (PBE+U, r2SCAN); 0.12--0.25 (HSE06); 0.50 (BHandHLYP) & $\leq$0.04 (hybrids, HF); $\leq$0.12 (others) & $+$0.06--0.07 (hybrids); $\leq$0.01 (HF); $\leq$0.24 (others) \\
magnetic reference & broken symmetry / high spin (HSE06, state a$'$) & one pair O reverses & 0.005--0.016 & 0.067 (0.071 $\to$ 0.004) \\
Mn pattern ($x$ = 0.5) & cluster-lowest / periodic & up to 0.7 (HSE06); cluster-lowest is an artifact & -- & -- \\
\midrule
\multicolumn{5}{l}{\emph{state search (observed failure)}} \\
missed lower state & state a$'$ $\to$ a (HSE06, BHandHLYP) & HSE06: 0.09--0.20 (R1, R2); BHandHLYP: up to 0.45 (R1) & $\leq$0.06 & $-$0.07 to $-$0.08 (HSE06); $-$0.01 to $-$0.03 (BHandHLYP) \\
\bottomrule
\end{tabularx}

\raggedright\footnotesize\textsuperscript{\emph{a}}Reference: R2, NO threshold 0.002, 3.6\,\AA{}, cfg07, broken symmetry, lowest stable state. Spins in e; the other columns are dimensionless. \textsuperscript{\emph{b}}PBE+U, $U$ = 5.0\,eV: $p$ = 2.29 lies 0.04 outside the strict interval, and the far option ($2M_S$ = 3, 0.71 from $p$) moves one oxygen to $-0.53$.
\end{table}

\subsection{A sample-based quantum solver on the embedded Hamiltonian}\label{sec:sqd}

A quantum solver would replace DMRG inside the embedding. SQD\cite{sqd2025} has recently been run on hardware for an open-shell 3d complex in implicit solvent\cite{sqdco2026} and combined with density-matrix embedding for molecules,\cite{dmetsqd,mlqscidmet2026} but neither setting has a spin-polarized embedding environment, the feature that shapes the Hamiltonian here. We test SQD on the HSE06 R2 problem in state a$'$ (Section~\ref{sec:sqddet}). This stable state lies 2.24\,\mEh{} above the lowest state found by the extended search (Section~\ref{sec:location}); the conclusions of this section concern the embedded Hamiltonian and the solver, which do not depend on that choice. DMRG in the full R2 space (25 MP2 natural orbitals) gives pair spins of $-0.806$/$-0.904$ (L\"owdin). The UNO spectrum is clean: 8 open-shell UNOs with occupations 0.67--1.21 (Mn 3d and the pair O 2p), 4 O 2p bonding orbitals at 1.96--1.99, and correlating orbitals from 0.03 down. The 12-, 14- and 16-orbital reduced active spaces take the 8 open-shell orbitals together with 3, 4 and 4 bonding and 1, 2 and 4 correlating ones; the remaining UNOs are frozen (doubly occupied) or dropped. This setup separates the embedding-side error, which a device cannot fix (exact diagonalization in the reduced active space vs DMRG in the full space), from the solver error, which a device would have to beat (SQD vs exact diagonalization of the same reduced-active-space Hamiltonian). At 12--16 orbitals exact diagonalization is routine ($3\times10^6$ determinants at 14 orbitals, $6\times10^7$ at 16), so the test demonstrates the correctness of the pipeline from embedding to an unrestricted qubit Hamiltonian and to SQD; it makes no claim about quantum resources or advantage.

\begin{figure}[tbp]
\centering
\includegraphics[width=\textwidth]{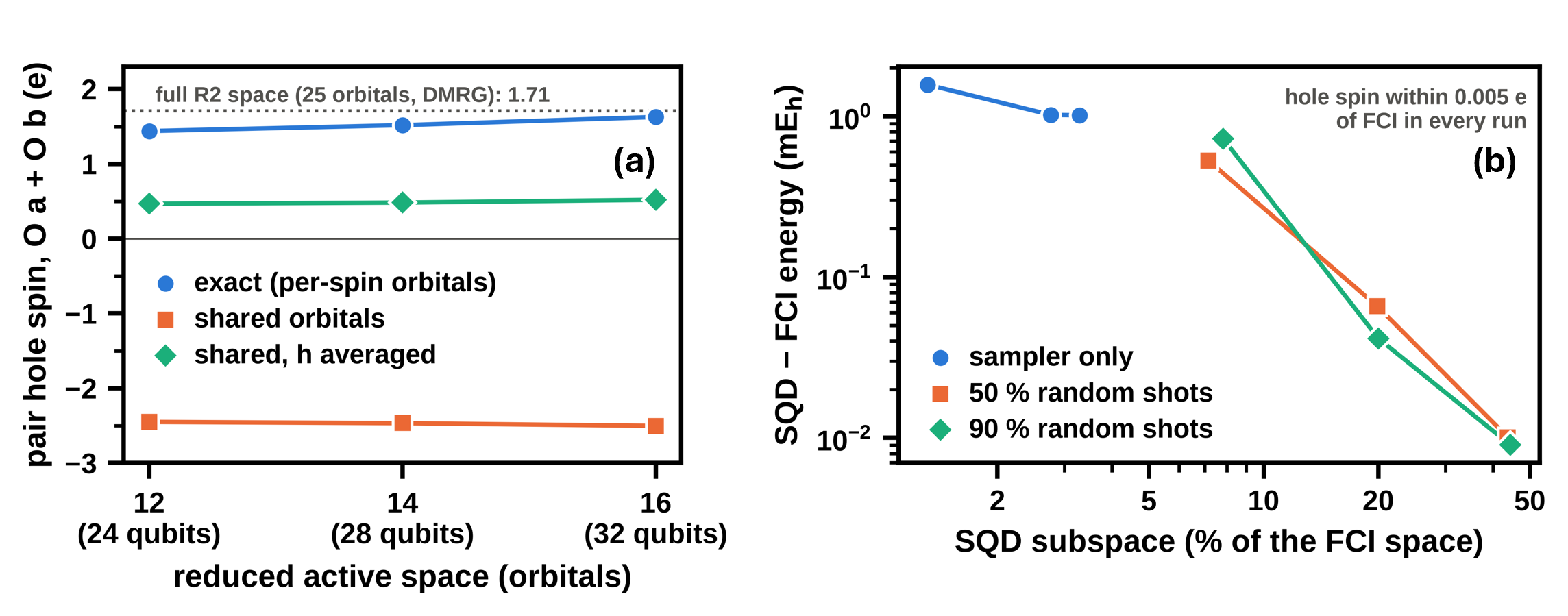}
\caption{SQD on the embedded R2 Hamiltonian (HSE06, cfg07). (a) Summed hole spin on the O pair, $-(s_a + s_b)$, from exact diagonalization in 12-, 14- and 16-orbital reduced active spaces (24--32 qubits) for the exact spin-unrestricted Hamiltonian (per-spin orbitals), the shared-orbital approximation, and the shared approximation with a spin-averaged one-body term (the input of standard SQD codes); dotted line, DMRG in the full 25-orbital R2 space. (b) SQD energy error on the exact 14-orbital Hamiltonian against the size of the recovered subspace, for samples from the UCJ circuit alone and with 50\,\% and 90\,\% of the shots replaced by random bitstrings.}
\label{fig:sqd}
\end{figure}

\paragraph{The embedded Hamiltonian is spin-unrestricted.} With a spin-polarized environment, projection embedding gives each spin its own environment and therefore its own active-subsystem orbital space. In R2 the $\alpha$ and $\beta$ environments share 125 of their 132 orbitals; the rest are environment Mn 3d orbitals and O 2p orbitals with spin-dependent Mn 3d tails, which reach into the region, so that every UNO of the reduced active space overlaps each spin's environment by 0.5--13\,\%. Projecting each spin's reduced active space off its own environment and L\"owdin-orthonormalizing it (overlaps with the UNOs $\geq$0.93 for $\alpha$ and $\geq$0.83 for $\beta$) yields $\alpha$ and $\beta$ reduced-active-space orbitals that differ ($\langle\phi_\alpha|\phi_\beta\rangle$ down to 0.75), so that $h^\alpha \neq h^\beta$ and the three two-electron blocks of eq~\ref{eq:uham} differ. Using the same UNOs for both spins, with one two-electron tensor and $h^\alpha \neq h^\beta$, places 0.5 $\alpha$ and 0.9 $\beta$ electrons of the reference in their own spin's frozen environment, violating the Pauli constraint that the projector enforces; averaging $h$ over spins, as required by the standard SQD input, additionally removes the spin polarization. The consequences are qualitative (Table~\ref{tab:window}, Figure~\ref{fig:sqd}a): the shared-orbital Hamiltonian reverses the hole spin and the averaged one nearly removes it. The 14- and 16-orbital reduced active spaces give the same picture (exact $-0.691$/$-0.828$ and $-0.748$/$-0.880$; shared $+1.317$/$+1.150$ and $+1.337$/$+1.166$; averaged $-0.173$/$-0.310$ and $-0.205$/$-0.315$). We therefore keep the unrestricted Hamiltonian and give SQD the unrestricted subspace solver of Section~\ref{sec:sqddet}. Any embedding with an open-shell environment needs this, and closed-shell test systems such as water and propionitrile cannot reveal it.

\begin{table}[tbp]
\caption{Exact Diagonalization in the 12-Orbital Reduced Active Space for Three Forms of the Embedded Hamiltonian}
\label{tab:window}
\centering
\begin{tabular}{lccc}
\toprule
Hamiltonian & $\langle S^2 \rangle$ & pair-O spin (L\"owdin) & pair-O spin (IAO) \\
\midrule
exact (per-spin orbitals) & 7.65 & $-0.665$ / $-0.775$ & $-0.646$ / $-0.768$ \\
shared orbitals & 10.62 & $+1.311$ / $+1.139$ & $+1.382$ / $+1.182$ \\
shared orbitals, $h$ averaged\textsuperscript{\emph{a}} & 6.00 & $-0.156$ / $-0.311$ & $-0.116$ / $-0.291$ \\
\bottomrule
\end{tabular}

\raggedright\footnotesize\textsuperscript{\emph{a}}The input form of standard SQD codes (one one-body matrix and one two-electron tensor).
\end{table}

\paragraph{Truncation to a device-sized active space.} Exact diagonalization of the 12-orbital reduced active space recovers 0.13--0.14\,e less hole spin on each pair oxygen than DMRG in the full R2 space ($-0.665$/$-0.775$ against $-0.806$/$-0.904$). The gap is 0.08--0.12\,e at 14 orbitals and 0.02--0.06\,e at 16 orbitals. At 12 orbitals the truncation is as large as the functional spread and the boundary steps of Section~\ref{sec:boundary}; at 32 qubits it falls below them. Two mean-field references put these numbers in context. The reference determinant of the reduced active space (the lowest $n_\alpha$ and $n_\beta$ UNOs, the starting point of the sampler) puts $+0.23$/$+0.13$ on the pair oxygens, the wrong sign, and lies 0.53--0.57\,\Eh{} above FCI in every reduced active space; in this basis the hole is a correlated feature that the solver must recover. The embedded UHF itself (HF-in-HSE06 over the whole active-subsystem orbital space), however, gives $-0.820$/$-0.914$, within 0.015\,e of the full-space DMRG. In the full embedded space, correlation inside R2 hardly moves the hole, and this holds for every production run (Section~\ref{sec:levels}): its functional dependence enters through the embedding potential and the boundary, not through correlation in the active region. The hole spin is therefore a weak test of the correlated solver, and we use the energy as the stringent one.

\paragraph{Solver error.} The UCJ circuit is deliberately imperfect: UCCSD cannot be run on the unrestricted Hamiltonian with standard codes, so its amplitudes come from the shared-orbital Hamiltonian, which is qualitatively wrong and on which UCCSD does not converge. Its overlap with the exact ground state is 0.63 (12 orbitals) and 0.69 (14 orbitals). The random-bitstring noise model is favorable, because configuration recovery is designed to repair this kind of noise; correlated hardware noise may behave differently, and the results do not transfer to hardware without testing. In the 12-orbital reduced active space with the clean sampler ($p$ = 0), SQD on the exact Hamiltonian comes within 0.16\,\mEh{} of FCI using 8.5\,\% of the determinant space and within 0.10\,\mEh{} at 13\,\%, with hole spins equal to FCI to $5\times10^{-4}$\,e. With $p$ = 0.5 or 0.9 the error falls to 0.002--0.012\,\mEh{}, but the random shots then fill 33--90\,\% of the space, so the 14-orbital reduced active space is the informative test (Table~\ref{tab:sqd14}, Figure~\ref{fig:sqd}b). There, with the clean sampler, the recovered subspace saturates at 3\,\% (the support of the UCJ state) and the energy at $+1.0$\,\mEh{}. The hole spin is within $5\times10^{-3}$\,e of FCI already at 1.3\,\% of the space, but, as shown above, this is the easy part. The energy needs a broader subspace: with random shots it reaches 0.5--0.7\,\mEh{} at 7--8\,\%, 0.04--0.07\,\mEh{} at 20\,\% and 0.01\,\mEh{} at 45\,\%. Random strings alone are 130--530\,\mEh{} above FCI at the same subspace sizes (Table~\ref{tab:sqd14}), so the random shots improve the energy only by enlarging a subspace that the UCJ samples and configuration recovery have already anchored on the important strings, not through the random strings themselves. A classical selected configuration interaction on the same Hamiltonian and at the same subspace size is more accurate than SQD at every size in the 14-orbital space, by a factor of 2--10 (0.91, 0.21, 0.074, 0.011 and 0.001\,\mEh{} at 1.3, 3.3, 7.2, 20 and 44\,\%), and matches the FCI-weight selection; in the 12-orbital space it is comparable to SQD (0.19 and 0.042\,\mEh{} at 8.5 and 13\,\%, against 0.16 and 0.10\,\mEh{} for SQD; Table~S13). SQD on the averaged Hamiltonian converges to that Hamiltonian's own, wrong hole ($-0.17$ to $-0.38$ / $-0.31$ to $-0.61$).

\begin{table}[tbp]
\caption{SQD on the Exact 14-Orbital Embedded Hamiltonian, with Classical Baselines at the Same Subspace Size\textsuperscript{\emph{a}}}
\label{tab:sqd14}
\centering
\footnotesize\setlength\tabcolsep{4pt}
\begin{tabular}{ccccccc}
\toprule
samples/batch & $p$ & $\Delta E$, SQD & subspace & pair-O spin, SQD & $\Delta E$, SCI\textsuperscript{\emph{b}} & $\Delta E$, random\textsuperscript{\emph{c}} \\
\midrule
300  & 0   & $+1.57$ & 1.3\,\% & $-0.691$ / $-0.823$ & $+0.91$ & 482--517 \\
1000 & 0   & $+1.02$ & 2.8\,\% & $-0.691$ / $-0.824$ & $+0.29$ & 430--494 \\
3000 & 0   & $+1.01$ & 3.3\,\% & $-0.690$ / $-0.824$ & $+0.21$ & 362--502 \\
300  & 0.5 & $+0.53$ & 7.2\,\% & $-0.691$ / $-0.828$ & $+0.074$ & 360--478 \\
300  & 0.9 & $+0.72$ & 7.8\,\% & $-0.691$ / $-0.827$ & -- & -- \\
1000 & 0.9 & $+0.04$ & 20\,\%  & $-0.691$ / $-0.828$ & $+0.011$ & 248--370 \\
3000 & 0.9 & $+0.01$ & 45\,\%  & $-0.691$ / $-0.828$ & $+0.001$\textsuperscript{\emph{d}} & 129--287\textsuperscript{\emph{d}} \\
FCI  & --  & 0       & 100\,\% & $-0.691$ / $-0.828$ & 0 & 0 \\
\bottomrule
\end{tabular}

\raggedright\footnotesize\textsuperscript{\emph{a}}Energies relative to FCI in \mEh{}; subspace as a fraction of the FCI space; L\"owdin pair-O spins; $p$ is the fraction of shots replaced by random bitstrings. A subset of Figure~\ref{fig:sqd}b; all runs are in Table~S5. \textsuperscript{\emph{b}}Classical selected configuration interaction at the same subspace dimension (Section~\ref{sec:sqddet}; Table~S13). \textsuperscript{\emph{c}}Random $\alpha$ and $\beta$ strings at the same dimension, three seeds. \textsuperscript{\emph{d}}At 44\,\%.
\end{table}

For quantum-hardware runs on embedded cathode models, three conclusions follow. The embedded Hamiltonian must be built spin-unrestricted when the environment is spin-polarized, and the environment leak of any shared basis should be checked (here 0.5--0.9 electrons, enough to reverse the hole). The solver is not the limiting error at this size: SQD reproduces the FCI hole spin of the reduced active space ($\leq 5\times10^{-3}$\,e) from 1--10\,\% of the determinant space and its energy to $\leq$0.07\,\mEh{} from about 20\,\%, although a classical selected configuration interaction of the same size does as well or better, so the test establishes the correctness of the pipeline and not an advantage of the quantum sampler; while the active-space truncation ($\sim$0.1\,e) and the embedding settings of Sections~\ref{sec:location}--\ref{sec:geometry} (0.1--0.3\,e) dominate. And because the hole is largely fixed at the embedded mean-field level in this system, a solver should be judged by its energy, whereas energies as physical observables are not converged even at 30--40 orbitals (Section~\ref{sec:energies}).

\subsection{What the correlated solver adds, and the level at which each conclusion holds}\label{sec:levels}

\paragraph{Embedded mean field and DMRG.} Every production DMRG run starts from an embedded UHF solution, and comparing the two for all 82 runs (seven functionals, R1--R3, both geometries, the states a and a$'$, the other-O state and the high-spin reference; Table~S12) separates what the embedding fixes from what correlation adds. The pair-O spins differ by a median of 0.024, 0.037 and 0.011\,e in R1, R2 and R3 (0.025, 0.018 and 0.012\,e for the lowest states alone), no pair-O spin changes sign, and the coupling pattern of the pair is the same in every case, including the high-spin reference ($+0.87$/$-1.01$ with UHF, $+0.73$/$-0.99$ with DMRG). The hole location, its dependence on the functional and the geometry, the spin-count intervals (0.21--0.29\,e in R1 for PBE+U and r2SCAN with the embedded UHF, against 0.14--0.26\,e with DMRG), the effect of the missed states and the large PBE+U boundary steps are therefore all present at the embedded mean-field level. Correlation changes the hole systematically but modestly: it reduces the pair-O spin on 90\,\% of the oxygens (by 0.03\,e on average) and raises the O-2p share of the fractionally occupied natural orbitals (by a median of 0.05 in R2). Its largest effect is on the O--O covalency, which DMRG roughly doubles where the hole is partial or asymmetric (Mayer bond order 0.08 $\to$ 0.14 for PBE+U at $U$ = 3.0 and 3.9\,eV with the 33/28 count and 0.025 $\to$ 0.065 for BHandHLYP, both in R2 at the HSE06 geometry), whereas the covalency of HSE06 is already present in the mean field (0.063--0.070 against 0.069--0.072). The largest change of a spin, 0.24\,e, occurs for PBE+U at $U$ = 5.0\,eV in R3 at the HSE06 geometry, where DMRG makes the pair asymmetric ($-0.66$/$-0.98$ against $-0.90$/$-0.81$) and removes the covalency of the mean field (0.025 against 0.163). The correlated natural orbitals of R2 show static correlation in the Mn--O coupling of the hole (occupations 1.26, 1.10, 0.89 and 0.73 for the HSE06 state a), but non-spin-adapted DMRG on UHF orbitals does not restore spin purity: $\langle S^2 \rangle$ changes by a median of 0.07 from its UHF value (10.83 against 8.75 for a pure $S$ = 5/2 at $2M_S$ = 5), so the correlated description of the Mn--O magnetic coupling remains broken-symmetry. The central finding is thus that the embedding (environment functional, geometry, boundary and electronic state) determines the hole, and the correlated solver refines its covalency.

\paragraph{Level of evidence.} Table~\ref{tab:levels} lists the level at which each conclusion is established. The location of the hole and the absence of an O--O bond rest on the lowest stable SCF state and are consistent with, but not established energetically by, the correlated calculations, whose total energies are not converged. Statements about how the hole is shared are established at both levels, because the hole populations are converged and agree between the embedded mean field and DMRG. The energetic ordering of competing states remains unresolved.

\begin{table}[tbp]
\caption{Level at Which Each Conclusion Is Established}
\label{tab:levels}
\centering
\footnotesize
\begin{tabularx}{\textwidth}{>{\raggedright\arraybackslash}p{3.6cm} >{\raggedright\arraybackslash}X >{\raggedright\arraybackslash}X >{\raggedright\arraybackslash}p{2.4cm}}
\toprule
conclusion & embedded SCF / mean field & DMRG-in-DFT & status \\
\midrule
hole on the O pair & lowest stable state for every functional, cell and geometry & consistent; energies not converged & SCF level \\
no O--O bond & Mayer bond order $\leq$0.24 & same & both levels \\
hole sharing set by functional and geometry & present in the embedded UHF & pair-O spins within a median 0.02--0.04\,e & both levels \\
spin-count interval & 0.21--0.29\,e (R1) & 0.14--0.26\,e (R1), $\leq$0.09\,e (R2) & both levels \\
weak O--O covalency & present (HSE06) or about half (partial holes) & roughly doubled where the hole is partial & correlated level \\
missed lower states & SCF energies 2.2 and 8.0\,\mEh{} lower & up to 0.45\,e change & SCF level \\
other-O vs O-pair ranking & other-O survives only with HSE06 & $+2.7$ to $-3.2$\,\kcal{}, drifts with the NO threshold & unresolved \\
configuration and state energies & -- & not converged at 30--40 orbitals & unresolved \\
\bottomrule
\end{tabularx}
\end{table}

%=====================================================================
\section{Conclusions}\label{sec:conclusions}

We developed an open-shell (spin-unrestricted) implementation of projection-based embedding, validated it to the \uEh{} level for closed-shell, broken-symmetry and $n_\alpha \neq n_\beta$ references with point charges and DFT+U, and used it to ask how the choice of environment functional, together with the embedding boundary, decides where the anion-redox hole of delithiated \LiMnO{} ends up and how it is described.

Where the hole sits is robust. In every setting examined (seven environment functionals from 0 to 100\,\% exact exchange and $U$ = 3--5\,eV, both delithiation states, the PBE+U and HSE06 geometries with matched clusters, 3.6 and 4.5\,\AA{} clusters, and all three active regions) the lowest SCF state places the hole on the O pair of the periodic calculation, and the extended state search confirms this for the cases it covered. No O--O bond forms anywhere (Mayer bond order $\leq$0.24, against $\sim$1 for a peroxo bond). The one competing location, a hole on a different oxygen, survives as an SCF solution only with HSE06; at the correlated level its energy relative to the O pair is not resolved: it drifts from $+2.7$ to $-3.2$\,\kcal{} with the virtual space when the nonadditive exchange correction is included, and from $+10.1$ to $+4.1$\,\kcal{} without it. The contribution of the correction is stable along each series but changes from $+34$ to $-7$\,\kcal{} between two O-pair states that differ by 1.4\,\kcal{} at the SCF level.

How the hole is shared is set by the functional together with the geometry. At the PBE+U geometry the functionals split into two groups: DFT+U and r2SCAN give a partial hole (0.79--0.91\,e per pair oxygen in R2), while the hybrids and HF give a full, uncoupled one (0.99--1.04\,e, Mayer bond order $\leq$0.005). At the HSE06 geometry, where the pair contracts from 2.58 to 2.41\,\AA{}, the hybrids return to a partial (HSE06) or asymmetric (BHandHLYP) hole with weak O--O covalency (about 0.07), a change of up to 0.5\,e on one oxygen; only HF is insensitive to the geometry. For DFT+U and HSE06 at their own geometries the spin spreads across the active/environment boundary, so the integer \Ms{} of the active space becomes a choice. That choice moves a pair oxygen by 0.14--0.26\,e when the active region holds only the O pair, but by $\leq$0.09\,e in R2, which adds the Mn neighbors, making R2 the smallest region with a reliable boundary (in R3 the choice can again reach 0.23--0.36\,e). Weak O--O covalency is converged with region size only for HSE06 at its own geometry; elsewhere it is either about zero or decreases as the region grows. The virtual space, bond dimension, cluster size, boundary ECPs and solver are converged for the hole character in R1 and R2, and embedding agrees with full-cluster DMRG to $\leq$0.04\,e (R1) and $\leq$0.013\,e (R2). The embedded mean field already reproduces these hole descriptions: across all 82 production runs DMRG changes a pair-oxygen spin by a median of 0.02--0.04\,e (at most 0.24\,e) and never reverses it, and what the correlated solver mainly adds is O--O covalency, which it roughly doubles where the hole is partial or asymmetric. Total energies, and therefore configuration energies and state rankings, are not converged at feasible active spaces of 30--40 orbitals.

Several lessons apply beyond this system. First, state control is part of the method: an extended search with an ECP anneal and cross-seeding found stable states 2.2 and 8.0\,\mEh{} below the initially recorded HSE06 and BHandHLYP states, which changed a pair-oxygen spin by up to 0.45\,e and removed the O--O covalency; neither route alone found both, and unseeded SCFs, discarded unstable solutions and clusters re-carved at a new geometry produced apparent effects of 0.3--0.7\,e before they were controlled (Supporting Information, Pitfalls). Second, the geometry must be consistent with the environment functional, because a hybrid evaluated at a DFT+U geometry describes a different hole from the same hybrid at its own geometry, and compared clusters must be carved from the same atoms. Third, in open-shell embedding with a spin-unrestricted solver the per-spin orbital counts fix \Ms{}; when the spin of the region is not an integer, both bracketing values should be computed and reported as an interval, and the active region should include the metal neighbors of the hole-bearing ligands. Fourth, the cluster's lowest magnetic state should be checked against the periodic crystal, which at $x$ = 0.5 places the cluster-preferred Mn pattern 0.26\,eV higher. Fifth, rankings of competing open-shell states at feasible active-space sizes should be reported with and without the nonadditive exchange correction and treated as unresolved when the two disagree. Finally, correlating the level-shifted environment orbitals moves the CCSD(T)-in-SCAN spin gap of [Fe(H$_2$O)$_6$]$^{3+}$ by 4.4\,\kcal{}, so the correlation-space convention should be stated with every embedded spin gap.

For quantum solvers, the embedded Hamiltonian of a spin-polarized environment is intrinsically spin-unrestricted. In our R2 reduced active space the $\alpha$ and $\beta$ orbitals overlap down to 0.75, and a shared-orbital qubit Hamiltonian violates the Pauli constraint against the frozen environment by 0.5--0.9 electrons and reverses the hole spin. The unrestricted embedded Hamiltonian is nevertheless an ordinary qubit Hamiltonian, and SQD with an unrestricted subspace solver recovers the FCI hole character of the reduced active space from 1--10\,\% of the determinant space and its energy to $\leq$0.07\,\mEh{} from about 20\,\%, under a favorable noise model, whereas a classical selected configuration interaction of the same size is as accurate or more so (2--10 times in energy at 14 orbitals); this is a test of the pipeline, not of a quantum advantage. These tests used the HSE06 state a$'$, which the extended search later found to lie 2.24\,\mEh{} above the lowest state; they concern the Hamiltonian and the solver, not the state. A hardware run on this problem would be limited by the embedding (state search, geometry, boundary spin), not by the solver, and because the embedded mean field already gives the hole to within 0.015\,e, solver benchmarks should use the energy. Removing the boundary-spin interval requires either regions large enough for an integer spin at their edge or a spin-adapted partition. The weak O--O covalency of the geometry-consistent hybrid, about 0.07 and converged with region size, is the natural target for spectroscopic comparison, for example by O K-edge RIXS\cite{loaiza2026} or EELS.\cite{kunitsa2025}

%=====================================================================
\begin{suppinfo}
Literature search (S1); the Nbed spin-polarization error and its correction (S2); DFT-in-DFT exactness of the cluster partitions (S3); water validation tables; the Fe benchmark and the propionitrile reproduction in full; pitfalls met during production (P1--P9); description of the embedding code and the unrestricted SQD subspace solver (S7); the extended state search (S8); run-level Tables S1--S6 and S10--S13, including the embedded mean field of every production run (S12) and the classical baselines of the SQD tests (S13).
\end{suppinfo}

\section*{Data and Software Availability}
The embedding code is available at \url{https://github.com/pri-tam-das/pembed}. Input files, cluster geometries, point charges, run-level outputs, and the scripts that generate all tables and figures are available at \url{https://doi.org/10.5281/zenodo.23183011}. All calculations used open-source software (PySCF, block2, ffsim, qiskit-addon-sqd), except the periodic calculations, which used the licensed VASP code.
% CHECK before submission: the GitHub repository is public and the Zenodo record is published (not only reserved).

\begin{acknowledgement}
This work was supported by the Korea Research Institute of Chemical Technology (KRICT) (DK2685-01). During preparation of this manuscript, the authors used Claude for editorial assistance, including improving clarity and readability of selected text, and for assistance drafting and troubleshooting Python code used in data analysis.
\end{acknowledgement}

\bibliography{refs.bib}

\end{document}